\documentstyle[preprint,eqsecnum,pra,amssymb,aps]{revtex}
\input{epsf.sty}
\def\framegraphics{\def\ifframe{\iftrue}}
\def\dontframegraphics{\def\ifframe{\iffalse}}
\def\drawgraphics{\def\ifdraw{\iftrue}}
\def\dontdrawgraphics{\def\ifdraw{\iffalse}}

\dontframegraphics
\drawgraphics

\newcommand{\graphics}[6]{
\def\epsfsize##1##2{#6##1}
\begin{picture}(#2,#3)
  \ifframe
    \put(0,0){\framebox(#2,#3){}}
  \fi
  \ifdraw
    \put(0,#3){\begin{picture}(0,0)
                 \put(#4,#5){\epsfbox{#1}}
               \end{picture}}
  \fi
\end{picture}}
\tighten
\begin{document}
\draft
\preprint{ADP-93-208/M19}
\title{Phase-Space Decoherence: a comparison between
Consistent Histories and\\ Environment Induced Superselection}
\author{J. Twamley\cite{Author}}
\address{
Department of Physics and Mathematical Physics, University of Adelaide\\
GPO Box 498, Adelaide, South Australia 5001}
\date{\today}
\maketitle
\begin{abstract}
We examine the decoherence properties of a quantum open system as
modeled by a quantum optical system in the Markov regime. We look
for decoherence in  both the Environment Induced Superselection (EIS)
and Consistent Histories (CH) frameworks. We propose a general measure of the
coherence of the reduced density matrix and find that EIS decoherence
occurs in a number of bases for this model. The degree of
``diagonality'' achieved increases with bath temperature. We evaluate
the Decoherence Functional of Consistent Histories for coarse grained
phase space two-time projected histories. Using the measures proposed
by Dowker and Halliwell we find that the consistency of the histories
improves with increasing bath temperature, time and final grain size
and decreases with initial grain size. The peaking increases with
increasing grain size and decreases with increasing bath temperature.
Adopting the above proposed measure of  ``coherence'' to the Decoherence
Functional gives similar results. The results agree in general with
expectations while the anomalous dependence of the consistency on the
initial grain size is discussed.
\end{abstract}
\narrowtext

\section{Introduction}
The emergence of classical dynamics from the underlying quantum
behaviour still continues to be one of the foremost questions in
physics today. Two ingredients essential to this transition are (i)
the establishing of suitable ``classical'' correlations between
appropriate coordinates and their conjugate momenta (ie. classical
equations of motion)  and (ii) the suppression or destruction of 
quantum correlations. The need for the former is self-evident while
the latter is necessary to explain the general absence of macroscopic 
quantum superpositions. A paradigm which primarily addresses (ii) and
which attempts to describe this transition is {\em Decoherence}. In
this school of thought, Decoherence is the general term which
describes a dynamical process whereby the quantum correlations in the
system in question are
diminished or destroyed. These processes can be split into those which
deviate from ordinary quantum mechanics, eg. GRW theory and quantum
mechanical spontaneous localisation QMSL
\cite{GHIRARDI:1986}, and those processes which stay within
the confines of quantum mechanics. We will only consider the latter in
this paper.  The two primary paradigms which attempt to describe the
transition from quantum to classical using pure quantum mechanics 
are Environmental
Induced Superselection (EIS) and Consistent Histories (CH)
\cite{ZUREK:1991,HARTLE:1993a}.
Although much has been written concerning these two theories,
calculations comparing the two in non-trivial models are lacking. Even
in relatively 
simple models the computations quickly become lengthy and difficult
to interpret. 
In this paper we examine the dynamics of a simple quantum optical
system from the viewpoints of both  EIS and CH. Upon comparison we
find that they generally agree, that is, if EIS occurs in a particular
basis then histories in that basis become more consistent.

In section II we derive the general solution to the quantum
optical master equation in the Markov regime for a system
comprising of an harmonic oscillator linearly coupled to an infinite
bath of harmonic oscillators with the usual quantum optical
approximations. We solve for the evolution of the reduced density
matrix of the system in the position, momentum, number and coherent
state  bases. In section III we discuss
EIS in these bases using a general measure of ``diagonality''
which quantifies the peaking of the density matrix about the diagonal.
This measure is sensitive to the coherence of the system state.
We solve for the general behaviour of this measure in the position
basis for an arbitrary
initial system state and examine the particular case of a mixture of
two coherent states. We find that EIS occurs to some extent 
in all of the above bases with 
$\rho_{reduced}$ becomming more ``diagonal'' with time and 
increasing bath temperature.

In section IV we calculate the decoherence functional of CH for two-time
projected coarse-grained histories on phase-space.
As in all previous calculations,  we use
quasi-projectors, that is, operators which are complete but are not
exclusive, to define the coarse-grained histories on phase-space. We
then compute the degree of consistency and peaking of the decoherence
functional using measures advanced by Dowker and Halliwell
\cite{DOWKER:1992}. We find that the consistency and peaking behave as
expected except for an anomalous {\em decrease} in consistency with
increased initial grain size. This behaviour is also found in all
previous calculations. We also adopt the above proposed measure for
diagonality in the EIS framework to quantify the ``coherence'' of the
decoherence functional. We find a similar dependence of this measure on
the model parameters to that displayed by the the  Dowker-Halliwell measure. 
In the conclusion we discuss the results of the EIS and CH
calculations and the anomalous dependence of the consistency in the CH
picture on the initial graining. We also discuss the legitimacy of
using the ``coherence'' measure in the CH picture. We argue that in
the case of a set of many alternatives, this ``coherence'' measure
gives a good indication that the probability sum rule is met to the
given degree. 

Before examining the dynamics of the quantum optical master equation
we first make a few remarks concerning both the EIS and CH scenarios.

\subsection{Environment Induced Superselection}
Environmentally Induced Superselection was primarily developed by
Zurek in the early 1980's \cite{ZUREK:1981} in relation to quantum
non-demolition measurements of gravitational wave antennas. Since then,
the theory has steadily evolved to encompass more complicated and
realistic models. Essentially, the system in question is coupled to a
large external bath via a particular interaction with Hamiltonian
$H_{int}$. If the interaction is such that the reduced density matrix of
the system becomes approximately diagonal in a particular basis in a
time much less than the relaxation time then we say EIS has occurred.
The process is aimed primarily at removing the quantum coherence from
the system. The system's reduced density matrix in this particular
basis is approximately diagonal and is given a purely ignorance
interpretation. That is, the system is now in one of the states on the
diagonal but we do not know which one. 

A number of calculations have
shown that in some cases EIS is very successful and can damp away the
quantum superpositions very quickly indeed \cite{JOOS:1985,ZUREK:1991}.
Other models have shown however, that the spectral properties  of
environment play a  significant role in determining the
decoherence time \cite{HU:1992,PAZ:1992}. In this paper we will work with the
standard ohmic environment. 

It is also the case that the initial conceptions of EIS have changed.
The concept of a pointer basis $|\Lambda\rangle$, 
the particular basis in which the
reduced density matrix becomes diagonal, strictly holds only in those
models for which the interaction Hamiltonian commutes with the
number operator of the pointer basis
$\tilde{\Lambda}|\Lambda\rangle=\Lambda|\Lambda\rangle$. For more
realistic cases and for the models treated in this paper, no exact
pointer basis exist. Work on a suitable generalization of the concept
of a pointer basis, that is, a system basis which is least effected by the
interaction with the bath, is currently being pursued
\cite{ZUREK:1993,ZUREK:1993a}. Also, there does not appear to be a general
concrete description concerning the  ``diagonality'' of a reduced
density matrix in a particular basis. Some measures have been advanced
for gaussian states. For particular 
non-gaussian states 
a  fringe visibility measure based on the appearance of
the Wigner function has been proposed \cite{PAZ:1992}.
This however, appears to quantify
the amount of interference in the state rather than the coherence. EIS
is the process whereby  the reduced state of the system 
deviates more and more from a description as a single ray in Hilbert space.
Such a process reduces the interferences present in the state. However, pure
states, with unitary evolution, can display little interference. 
Such processes do {\rm not} display EIS. With this in mind 
we propose and investigate a
measure of the ``diagonality'' of the state in a number of  bases in a
quantum optical model. We find that the coherence of the reduced
density matrix of the system decays and gives clear evidence of EIS in
a number of bases. However, the degree of diagonality depends on the
basis chosen. It also depends on the model parameters in an expected
manner, becoming more diagonal with increasing bath temperature. 

We
have not examined the attainment of (i) above in the EIS framework.
This is because there appears to be some dispute concerning the
quantification of classical correlations in the state. Previously,
some have examined the Wigner function of the state for peaks
\cite{HABIB:1990}. However, the Wigner function is not a true probability
distribution and does not correspond to the results of any conceivable
measurement on the system. The $Q$ function of quantum optics {\em
does} correspond to the result of a particular measurement scheme
\cite{BRAUNSTEIN:1991}. 
More generally, there exist true probability distribution functions
corresponding to more general measurements \cite{CAVES:1986}. 
To establish the existence and to quantify ``classical'' correlations
one should examine such distributions. Work towards this goal is in
progress \cite{ANDERSON:1993}.

Finally, the ignorance interpretation of the resulting diagonal reduced
density matrix seems part and parcel of EIS. This feature of the
framework has 
surprisingly not attracted much attention in the literature. If EIS is to
describe a truly ontological reality then Zurek's arguments may not suffice
\cite{DESPAGNAT:1990}. It is clearly capable of describing an
empirical reality \cite{DESPAGNAT:1990} but are it's aspirations
higher than just such a description?

\subsection{Consistent Histories}
In order not to confuse the reader with various different meanings of
the term  decoherence,
we will refrain from the more usual nomenclature of ``Decohering
Histories''
 and
use the more descriptive terminology --  ``Consistent Histories''. 
Consistent Histories was initially researched 
by Griffiths and Omn\'{e}s \cite{GRIFFITHS:1984,OMNES:1992} and later on 
independently rediscovered and greatly expanded on by Gell-Mann and
Hartle \cite{GELL-MANN:1990}. This paradigm is more ambitious than that
of EIS as it attempts to describe the emergence of the complete
quasi-classical world of familiar experience from the underlying
quantum world. Central to this framework is the notion of
coarse-grained histories, each with an assigned probability. A primary
goal is the identification of complete sets of coarse-grained
histories which exactly satisfy the probability sum rules. Although
much has been written concerning this theory, there are only a few
calculations on realistic models \cite{DOWKER:1992,HARTLE:1993a}. 
Quantum optics is an avenue where both quantum and
classical behaviour are accessible to the experimentalist. We will
in particular look at coarse-grained histories on phase-space itself. Previous
calculations have concentrated on position projected histories with
momenta information gained through time-of-flight. In this paper we
look for consistent histories in {\em both} position and momentum. 
We also note that the interpretation of the
diagonal decoherence functional is typically a relative state
interpretation and thus the interpretational problems concerning which history
is realized are moot (however see \cite{DESPAGNAT:1990}).

\section{Quantum Optical Master Equation}
We begin by looking at the optical master equation for a system
linearly coupled via the usual quantum optical coupling to a bath of
harmonic oscillators \cite{GARDINER:QUANTUM_NOISE,TWAMLEY:1993a}. For
completeness and to relate this to earlier work let us first consider
a system with a Hamiltonian
\begin{equation} H/\hbar=\omega (t)\,a^{\dagger}
a+f_{1}(t) \,a+f^{*}_{1}(t)\,a^{\dagger}+
f_{2}(t)\,a^{2}+f^{*}_{2}(t)\,a^{\dagger}{}^{2}\;\;.\label{eq0.1}
\end{equation}
In the quantum optical regime we can write the master equation for
the reduced density matrix of the system as
\cite{GARDINER:QUANTUM_NOISE}
\begin{equation}
\dot{\rho}=-\frac{i}{\hbar}[H,\rho]+\Lambda\rho={\cal L}\rho\;\;,
\label{eq1.1}\end{equation}
\begin{equation}
{\Lambda}\rho=
\frac{\gamma(\bar{n}+1)}{2}\left\{[a ,\rho a^{\dagger} ]+
[a\rho,a^{\dagger} ]\right\}
+\frac{\gamma\bar{n}}{2}\left\{[a^{\dagger} ,\rho a ]+[a^{\dagger}\rho
,a ]\right\}\;\;.\label{eq1.2}
\end{equation}
This may be converted into a partial differential equation for a
Quasi-Distribution Function (QDF), $W(\alpha ,\alpha^{*}, s)$ 
(or $W(\alpha,s)$ for short) where 
\cite{CAHILL:1969}
\begin{equation}
W(\alpha,\alpha^{*},s,t)=\frac{1}{\pi}\int\,
\chi(\xi,s,t)e^{\alpha\xi^{*}-\alpha^{*}\xi}\,d^{2}\xi\;\;,\label{eq2}
\end{equation}
\begin{equation}
\chi(\xi,s,t)={\rm Tr}[\rho(t) {\rm
D}(\xi,s)]\;\;,\label{eq3}\end{equation}
\begin{equation}
{\rm D}(\xi,s)=\exp (\xi a^{\dagger}-\xi^{*}a+\frac{1}{2}s|\xi
|^{2})\;\;,
\label{eq4}\end{equation}
where we have the normalization $\int W(\alpha,s,t)d^{2}\alpha/\pi=1$.
The family of distribution functions $W(\alpha,s,t)$ encompasses 
the more familiar distributions. For $s=1$, $W(\alpha,+1,t)=P(\alpha,t)$ is
the P distribution of Quantum Optics. For $s=0$,
$W(\alpha,0,t)=W(\alpha,t)$ is the familiar Wigner distribution while for
$s=-1$, $W(\alpha,-1,t)=Q(\alpha,t)$ is the Q or Husimi distribution. The
different distributions correspond to different operator orderings.
 From (\ref{eq1.1}) we can derive a partial differential equation for
both $W(\alpha,s)$ and $\chi(\xi,s)$ from first principles using
equations (\ref{eq2}, \ref{eq3}, \ref{eq4}). However, using the general rules
derived in Vogel and Risken \cite{VOGEL:1989} we can easily obtain
\widetext\begin{eqnarray}
&&\dot{W}(\alpha,s,t)= \nonumber\\
&&-\frac{i}{\hbar}\left\{
 [f_{1}+2\alpha f_{2}+\alpha^{*}\omega ]\partial_{\alpha^{*}}
-[f_{1}^{*}+2\alpha^{*}f_{2}^{*}+\alpha\omega]\partial_{\alpha}
 -s[f_{2}\partial^{2}_{\alpha\alpha^{*}}-f_{2}^{*}
\partial^{2}_{\alpha\alpha^{*}}]\right\}W(\alpha,s,t)\\
&&+\frac{\gamma}{2}\left\{\partial_{\alpha}\alpha+
\partial_{\alpha^{*}}\alpha^{*}
+[2\bar{n}+1-s]\partial^{2}_{\alpha\alpha^{*}}\right\}W(\alpha,s,t)
\;\;,\label{eq6}\end{eqnarray}
\begin{eqnarray}
&&\dot{\chi}(\xi,s,t)=\nonumber\\
&&-\frac{i}{\hbar}\left\{\omega[\xi^{*}\partial_{\xi^{*}}-\xi\partial_{\xi}]
-[\xi f_{1}+\xi^{*}f_{1}^{*}]
-s[\xi^{2}f_{2}-\xi^{*\,2}f_{2}^{*}]
+2[f_{2}\partial_{\xi^{*}}-f_{2}^{*}\partial_{\xi}]\right\}\chi(\xi,s,t)\\
&&+\frac{\gamma}{2}\left\{(s-1-2\bar{n})\xi\xi^{*}-\xi^{*}\partial_{\xi^{*}}
-\xi\partial_{\xi}\right\}\chi(\xi,s,t)\;\;,\label{eq7}\end{eqnarray}
\narrowtext
where we have dropped the $t$ dependence of $f_{i}$ and $\omega$.
We see that the PDE for $\chi$ is first order. The general solution
for this model was treated in \cite{TWAMLEY:1993a}. We will instead
specialize to the case of a single harmonic oscillator with constant
frequency linearly coupled to the bath oscillators. Going to the
interaction picture we get
\narrowtext
\begin{equation}\dot{W}(\alpha,s,t)=\frac{\gamma}{2}
\left\{\partial_{\alpha}\alpha
+\partial_{\alpha^{*}}\alpha^{*}+[2\bar{n}+1-s]\partial^{2}_{\alpha\alpha^{*}}
\right\}W(\alpha,s,t)\;\;,\label{eq8}\end{equation}
\begin{equation}\dot{\chi}(\xi,s,t)=-\frac{\gamma}{2}\left\{
(2\bar{n}+1-s)\xi\xi^{*}+\xi^{*}\partial_{\xi^{*}}+\xi\partial_{\xi}\right\}
\chi(\xi,s,t)\;\;.\label{eq9}\end{equation}
Using the method of characteristics we will solve for the dynamics of
this model for an initial state
$\rho=\sum_{\alpha,\beta}N_{\beta\alpha}|\alpha\rangle\langle\beta|$
where ${\rm
Tr}\rho=\sum_{\alpha\beta}N_{\beta\alpha}\langle\beta|\alpha\rangle=1$.
Thus $\rho$ is a general superposition of coherent states at $t=0$.
(Coherent states are eigenstates of the lowering operator of the
oscillator algebra i.e. $a|\alpha\rangle=\alpha|\alpha\rangle$). Since
the coherent states are overcomplete any initial state may be
represented thus. Let us solve (\ref{eq9}) for $\chi(\xi,s=-1,t)$ when
$\rho (t=0)=|\alpha_{0}\rangle\langle\beta_{0}|$. From (\ref{eq3}) and
using $Q(\alpha)=\langle\alpha|\rho|\alpha\rangle = W(\alpha,-1)$ we get
\begin{equation}
\chi(\xi,-1,t)=\int\,\frac{d^{2}\alpha}{\pi}e^{\xi\alpha^{*}-\xi^{*}\alpha}\,
Q(\alpha,t)\;\;.\label{chi_def}\end{equation}
Using this we can calculate the form of the characteristic function
$\chi$ at
$t=0$ to be
\begin{equation}
\chi(\xi,-1,t=0)=\langle\beta_{0}|\alpha_{0}\rangle\exp 
[-|\xi|^{2}-\xi^{*}\alpha_{0}+\beta_{0}^{*}\xi]\;\;.\label{eq10}\end{equation}
For $s=-1$ we can rewrite (\ref{eq9}) as
\begin{equation}
\chi_{,t}+\frac{\gamma}{2}a\chi_{,a}+\frac{\gamma}{2}b\chi_{,b}
=-\gamma(1+\bar{n})ab\chi\;\;,
\label{eq12}\end{equation}
where $a=\xi$ and $b=\xi^{*}$ are now treated as independent
variables.
Looking at the characteristics of (\ref{eq12}) we see that
\begin{equation}
dt=\frac{da}{\gamma a/2}=\frac{db}{\gamma b/2}=
\frac{d\chi}{-\gamma (1+\bar{n})ab\chi}\;\;.
\label{eq13}\end{equation}
Integrating this we obtain three constants,
\begin{equation}
k_{1}=\frac{2}{\gamma }\ln a-t\;,\qquad
k_{2}=\frac{2}{\gamma }\ln b-t\;,\nonumber\end{equation}\begin{equation}
k_{3}=-\frac{\ln \chi}{1+\bar{n}}-ab\;\;.\label{eq13a}\end{equation}
Solving for $a,b$ and $\chi$ in terms of $k_{i}$ and inserting these
into (\ref{eq10}) at $t=0$ gives a relationship between the $k_{i}$
which is valid for $t\geq 0$. Re-inserting the $t$ dependence back into
the $k_{i}$ and going back to the Heisenberg picture yields
\begin{equation}
\chi(\xi,s=-1)=\langle\beta_{0}|\alpha_{0}\rangle
\exp
[-\kappa(t)|\xi|^{2}+\beta_{0}^{*}(t)\xi-\alpha_{0}(t)\xi^{*}]\;\;,
\label{eq14}\end{equation}
where $\kappa(t)=1+\bar{n}(1-e^{-\gamma t})$,
$\beta_{0}(t)=\beta_{0}e^{-(\gamma /2+i\omega)t}$ and
 $\alpha_{0}(t)=\alpha_{0}e^{-(\gamma /2+i\omega)t}$.
Inverting (\ref{chi_def}) to obtain the $Q$ distribution we get
\begin{eqnarray}
Q(\alpha,t) & = &
\int\,\chi(\xi,\xi^{*})\,e^{\alpha\xi^{*}-\alpha^{*}\xi}
\frac{d^{2}\xi}{\pi}\nonumber\\
 & = & \frac{\langle\beta_{0}|\alpha_{0}\rangle}{\kappa(t)}\exp\,
\left[-\frac{1}{\kappa(t)}(\alpha-\beta_{0}(t))^{*}
(\alpha-\alpha_{0}(t))\right]\;\;,\label{eq15}\end{eqnarray}
where we have used (\ref{appendix1}) from Appendix A.
To obtain the off-diagonal elements
$\langle\tilde{\alpha}|\rho(t)|\tilde{\beta}\rangle$ we use the
identity 
\begin{equation}\rho  =  \int\frac{d^{2}\xi}{\pi}
e^{-\xi a^{\dagger}}e^{\xi^{*} a} \chi(\xi,\xi^{*})\;\;,
\label{eq17}\end{equation}
and inserting (\ref{eq14}) we get
\begin{eqnarray}
&&\langle\tilde{\alpha}|\rho(t)|\tilde{\beta}\rangle=\nonumber\\
&&\frac{\langle\beta_{0}|\alpha_{0}\rangle}{\kappa(t)}
\exp\,\left[-\frac{1}{\kappa(t)}(\tilde{\alpha}-\beta_{0}(t))^{*}
(\tilde{\beta}-\alpha_{0}(t))\right]\langle\tilde{\alpha}|
\tilde{\beta}\rangle
\;\;.\label{eq18}\end{eqnarray}
Using equation (\ref{eq18}) we can now examine the
 general case where $\rho(t=0)=\sum_{\alpha_{0}\beta_{0}}\,
N_{\beta_{0}\alpha_{0}}|\alpha_{0}\rangle\langle\beta_{0}|$ 
and can evaluate $\rho(t)$ to be
\begin{eqnarray}
\rho(t) &=& \int\frac{d^2\alpha d^2\beta}{\pi^2}\,
\langle\alpha|\rho|\beta\rangle|
\alpha\rangle\langle\beta| \;\;,\nonumber\\
\langle\alpha|\rho(t)|\beta\rangle&=&\langle\alpha|e^{\int{\cal L}dt}\rho(0)
|\beta\rangle\;\;,\end{eqnarray} 
giving 
\widetext
\begin{equation}
\rho(t)=
\sum_{\alpha_{0}\beta_{0}}\frac{N_{\beta_{0}\alpha_{0}}}
{\kappa(t)}\langle\beta_{0}|\alpha_{0}\rangle
\int\,\frac{d^2\alpha d^2\beta}{\pi^2}
\langle\alpha|\beta\rangle
\exp\,
\left[
-\frac{1}{\kappa(t)}
(\alpha-\alpha_{0}(t))^{*}(\beta-\beta_{0}(t))
\right]\,
|\alpha\rangle\langle\beta|
\;\;,\label{eq20}\end{equation}
where ${\cal L}$ is the Markov super-operator (\ref{eq1.1}).
With this the dynamics is completely solved. However, to compare the decay of
off-diagonal elements of the reduced density matrix in other bases we will 
also compute $\rho$ in the position $x$, momentum $p$ and number 
basis.

\widetext
To compute the off-diagonal elements in the position basis we again
use equation (\ref{eq17}), giving
\begin{equation}
\langle x|\rho(t)|y\rangle=\langle x|\int\frac{d^2\xi}{\pi}
e^{-\xi
a^{\dagger}}e^{\xi^{*}a}\chi(\xi,t)|y\rangle\;\;,\label{xy}\end{equation}
where $\chi(\xi,t)$ is given in (\ref{eq14}). From Appendix B we have
\begin{equation}
\langle x|e^{-\xi a^{\dagger}}e^{\xi^{*}a}|y\rangle =
\exp\,\left[ \frac{|\xi|^2}{2}+\frac{\xi^{*\,2}-\xi^{2}}{4}+
x\sqrt{\frac{\omega}{2\hbar}}(\xi^{*}-\xi)\right]\,
\delta(\xi_{x}+\sqrt{\frac{\omega}{2\hbar}}(x-y))\sqrt{\frac{\omega}{2\hbar}}
\;\;.\label{xy1}\end{equation}
For
$\rho(t=0)=\sum_{\alpha_{0}\beta_{0}}\,N_{\beta_{0}\alpha_{0}}|
\alpha_{0}\rangle\langle\beta_{0}|$
we can perform the gaussian integrations in (\ref{xy}) and use the result
(\ref{xy1}) to finally get
\begin{eqnarray}
&&\langle x|\rho(t)|y\rangle = \nonumber\\ 
&& \sqrt{\frac{2}{\pi\sigma_{+}^{2}}}\sum_{\alpha_{0}\beta_{0}}
N_{\beta_{0}\alpha_{0}}\langle\beta_{0}|\alpha_{0}\rangle\times\nonumber\\
&&
\exp\,\left[
-\frac{1}{2\sigma_{-}^{2}}X_{-}^{2}+\frac{1}{\sqrt{\sigma_{-}\sigma_{+}}}
(\alpha_{0}(t)-\beta_{0}^{*}(t))X_{-}-\frac{1}{2\sigma_{+}^{2}}
(X_{+}-\sqrt{\sigma_{+}\sigma_{-}}
(\alpha_{0}(t)+\beta_{0}^{*}(t)))^{2}\right]\;\;,
\label{eq21}\end{eqnarray}
where
\begin{equation}
\sigma_{+}^{2}=\frac{2\hbar}{\omega}(2\kappa(t)-1)\;,\qquad\qquad
\sigma_{-}^{2}=\frac{2\hbar}{\omega(2\kappa(t)-1)}\;\;,\label{sigma_def}
\end{equation}
and $X_\pm=x\pm y$.
Using similar methods we can calculate the elements in the momentum
basis to be
\begin{eqnarray}
&&\langle p_1 | \rho(t) | p_2 \rangle  =  \nonumber \\
&&
\frac{1}{\hbar\omega}\sqrt{\frac{\tilde{\sigma}_{-}^{2}}{2\pi}}\sum_{\alpha_0
\beta_0}
N_{\beta_0 \alpha_0}\langle \beta_0 | \alpha_0 \rangle\nonumber\\
&&\exp\,\left[
-\frac{1}{2\tilde{\sigma}_{-}^{2}}P_{-}^2
+\frac{i}{\sqrt{\tilde{\sigma}_{-}\tilde{\sigma}_{+}}}
(\beta_{0}^{*}(t)+\alpha_{0}(t))P_{-} -\frac{1}{2\tilde{\sigma}_{+}}(
P_{+}-i\sqrt{\tilde{\sigma}_{-}\tilde{\sigma}_{+}}
(\alpha_{0}(t)-\beta_{0}^{*}(t)))^{2}\right]\;\;,
\label{eq22}\end{eqnarray}
where
\begin{equation}
\tilde{\sigma}_{+}^{2}=2\hbar\omega(2\kappa(t)-1)\;,\qquad\qquad
\tilde{\sigma}_{-}^{2}=\frac{2\hbar\omega}{2\kappa(t)-1}\;\;,\end{equation}
and $P_\pm=p_1\pm p_2$.
To calculate the elements of $\rho$ in the number basis is slightly
tedious.
Noting that
\begin{equation}
\langle n | \rho | m \rangle=
\int\frac{d^2\alpha d^2\beta}{\pi^2}\langle n| \alpha\rangle\langle
\alpha|\rho|\beta
\rangle\langle\beta| m\rangle\;,\qquad\qquad\langle n|\alpha\rangle=
e^{-|\alpha|^2/2}\alpha^n/\sqrt{n!}\;\;,\label{eq23}\end{equation}
and the identity
\begin{equation}
A^n=\int\frac{d^2\alpha}{\pi}\alpha^n \, e^{-|\alpha|^2+A\alpha^{*}}=
\int\frac{d^2\alpha}{\pi}\alpha^{*\,n} \, e^{-|\alpha|^2+A\alpha}\;\;,
\end{equation}
we find, with some work, that
\begin{eqnarray} 
\langle n | \rho(t) | m \rangle= & & \nonumber\\
& &\frac{1}{\kappa(t)}\sum_{\alpha_0 \beta_0}N_{\beta_0 \alpha_0}
\langle \beta_0|\alpha_0\rangle
e^{-\frac{1}{\kappa(t)}\alpha_{0}^{*}(t)\beta_{0}(t)}
\,\Pi^{nm}(\alpha_0,\beta_0,t)\left(\frac{\alpha_0^*(t)}
{\kappa(t)}\right)^m\left(\frac{\beta_0(t)}{\kappa(t)}\right)^n\;\;,
\label{eq24}\end{eqnarray}
where
\begin{equation}
\Pi^{nm}(\alpha_0,\beta_0,t)\equiv
\sum_{s=0}^{{\rm min}(n,m)}\frac{ n!m!}{s!(n-s)!(m-s)!}\left(
\frac{\kappa(t)(\kappa(t)-1)e^{\kappa(t)t}}{\beta_0\alpha_0^*}\right)^s\;\;.
\label{eq24a}\end{equation}
We note that even in the case where
$\rho(t=0)=|\alpha_0\rangle\langle\alpha_0|$ equations (\ref{eq24}) and
(\ref{eq24a}) do not reduce to a simple tractable forms. 
In what
follows we will not make much use of equations  (\ref{eq24}) and
(\ref{eq24a}) except to note that they become diagonal as $t\rightarrow\infty$.
Finally we note that the results for $\langle x|\rho| y\rangle$ found here 
agree with those found in \cite{SAVAGE:1985}.
\narrowtext

\section{Environment Induced Superselection}
We now wish to examine the behaviour of the off-diagonal elements of
$\rho$ in a number of different bases and to determine whether EIS
occurs. There have been a number of proposals for the measure 
of the degree of diagonality of the density matrix
\cite{MORIKAWA:1990,LAFLAMME:1993}. 
These have mostly been defined however for a
gaussian $\rho$. Here we expand these definitions to a measure
$\Omega_{\alpha}$ of the ``diagonality'' of a general density matrix
in a particular basis $\alpha$ ($\alpha$ here denotes a general basis
i.e. x, p, n or coherent state).

 From recent work 
\cite{ALBRECHT:1992,ALBRECHT:1991,ZUREK:1993a,ZUREK:1993} there appears to
exist some competition between the attainment of a stable predictable pointer
basis and the attainment of EIS type decoherence. 
In this paradigm one must be careful not to confuse the loss of
coherence with the loss of interference.
In this paper we interpret EIS as a process whereby the {\em coherence} of the
state is lost. 
 From Zurek's earlier work the signature
of decoherence was the vanishing of the {\em coefficients} of the
off-diagonal elements of the reduced density matrix of the system in a 
given basis. The
decay of the interference in a particular basis 
was due to the decay of these coefficients. However, the loss of
interference can also be brought about by
sending the overlap of the off-diagonal states in that basis to
zero. 
For example, in Young's slits,  when one sets the slit
separation to be very large in comparison to the screen distance, the
magnitude of the interference terms decays due to the vanishing of the
overlap of these off-diagonal elements in the position basis. 
(There may, of course, be significant overlap in other bases, eg. the momentum
basis.) In this case
the loss of interference 
is  {\em not} due to the decay of the coefficients associated
with these off-diagonal elements. The state remains pure, no
coherences have been destroyed yet the quasi-probability distribution
function {\em tends} towards (but is never exactly)
that displayed by a true probability
distribution. Recently decoherence in the EIS paradigm has been 
defined as the absence of
interference fringes in the Wigner function in the special case of a
superposition of two gaussians \cite{PAZ:1992}. 
These interference fringes decay in
magnitude for large separations of the two gaussians. They never
disappear except in the case of infinite separation. The fringes also
vanish dynamically with time if one introduces a coupling with an
infinite bath. In the latter case coherence has been lost and
the off-diagonal coefficients decay while in the former case the
interference terms vanish because the overlap of the off-diagonal elements 
in the chosen basis
goes to zero. Thus, while the loss of coherence necessitates the loss
of interference the loss of interference need not result in a loss of
coherence. In this paper we follow the spirit of earlier work and define
Environment Induced Superselection as the dynamical process whereby the
reduced state of the system deviates from a single ray in Hilbert
space and moves towards a description as an improper mixture
(for more information on improper mixtures see \cite{DESPAGNAT:CONCEPTUAL}). 

The most obvious measure of the deviation of a given state $\rho$ from
being a ray in Hilbert space is the statistical distance between the
state $\rho$ and the closest mixed state 
\cite{HUBNER:1992,BURES:1969,CAVES:1993}
\begin{equation}
D_{1}(\rho_{1})\equiv \min_{\rho_{m}\in {\rm diag}\,\rho}\,
d^{2}_{B}(\rho_1 ,\rho_{m})\;\;,\end{equation}
where
\begin{equation}
d_{B}^{2}(\rho_1,\rho_2)=2(1-{\rm Tr}\sqrt{
\rho_{1}^{1/2}\rho_{2}\rho_{1}^{1/2}})\;\;,\label{eq25}
\end{equation}
is the Bures metric on the state space of all pure and impure states
$\rho$ \cite{BURES:1969}. Equation (\ref{eq25}) is very difficult to
evaluate in general. Instead we will generalize a measure of coherence
for gaussian $\rho$ first 
introduced by Morikawa \cite{MORIKAWA:1990} and used later in
\cite{BRANDENBERGER:1990,LAFLAMME:1993}. For $\rho$ of the form
\begin{equation}
\langle \bar{x}|\rho|x\rangle =
D \exp\left[
-A(x-\bar{x})^2-B(x+\bar{x})^2+iC(x-\bar{x})(x+\bar{x})\right]\;\;,
\end{equation}  
Morikawa defined a measure of coherence to be
\begin{equation} {\rm QD}\equiv
\frac{A}{B}\;\;.\label{eq26}\end{equation}
When ${\rm QD}>1$ the magnitude of the elements of the density matrix
in the position basis are peaked preferentially along the diagonal
$x=\bar{x}$. For $\rho$ pure, ${\rm QD}=1$. The natural generalization
of this measure to non-gaussian $\rho$ is to compute the ratio of the
off-diagonal variance to the on-diagonal variance of the squared
magnitude of $\rho$ in a specified basis
$\hat{\eta}|\eta\rangle=\eta|\eta\rangle$, that is
\begin{equation}
\Omega_{\eta}\equiv\frac{\int\int\,d\eta_{1}\,d\eta_{2}\,
\langle\eta_1|\rho|\eta_2\rangle\langle\eta_2|\rho|\eta_1\rangle
|\eta_1-\eta_2|^{2}}
{\int\int\,d\eta_1d\eta_2\,\langle\eta_1|\rho|\eta_2\rangle
\langle\eta_2|\rho|\eta_1\rangle|\eta_1+\eta_2-\langle
\eta_1+\eta_2\rangle|^{2}}\;\;.\label{eq27}\end{equation}
For $\rho$ gaussian and $|\eta\rangle=|x\rangle$, $\Omega_{x}=1/{\rm
QD}$ from (\ref{eq26}). For pure states,
$\rho=|\psi\rangle\langle\psi|$ and
$\Omega_{\eta}=1\;\;\forall\,\eta$. This makes sense as 
no pure state undergoing unitary evolution can lose it's coherence.
The offset in the denominator gives the variance of
${\cal S}\equiv|\langle\eta_1|\rho|\eta_2\rangle|^2$ about the mean
$\langle\eta\rangle$. Distributions of $\cal S$ which are concentrated
along $\eta_{-}=\eta_1-\eta_2=0$ result in $\Omega_{\eta}<1$. As $\cal
S$ becomes more concentrated along $\eta_{-}=0$, $\Omega_{\eta}$
decreases to zero. From (\ref{eq27}) the value of $\Omega_{\eta}$
depends on the basis chosen. This we expect as some basis will
display a greater peaking about their diagonal than others.  The
advantage of the dimensionless measure (\ref{eq27}) is that one can
compare $\Omega_{\eta}$ and $\dot{\Omega}_{\eta}$ for various bases
$|\eta\rangle$. One can also consider bases which are discrete. In
this case the integrals in (\ref{eq27}) become discrete sums.
Heuristically the measure $\Omega_{\eta}$ tells us how damped the
off-diagonal coherences are in the basis $|\eta\rangle$. For pure
states, no matter what the configuration, these ``off-diagonal''
coherences never vanish. For every contribution to the ``on-diagonal''
elements of $\rho$ there are equal contributions to the
``off-diagonal'' elements. A classic example of this is the
superposition of two localized coherent states
$|\psi\rangle=|\alpha\rangle+|-\alpha\rangle$. For a pictorial
description of $\cal S$ for this state see Zurek \cite{ZUREK:1991}.
In this example the heights of the four peaks in Zurek's plots do not
change with increasing $|\alpha|^2$. No coherence is lost, however the
interference diminishes.

A quantification of the amount of interference present in a state is
generally given by the magnitude of a typical off-diagonal element.
For the special case of two superposed gaussians a measure of the
fringe visibility of the ripples of the Wigner function has been
proposed  \cite{PAZ:1992}. Another quantification of the
degree of ``off-diagonality'' would be
\begin{equation}
\hat{\Omega}_{\eta}\equiv\frac{1}{{\rm Tr}\,\rho^{2}}
\int\int\,d\eta_1 d\eta_2\,
\langle\eta_1|\rho|\eta_2\rangle\langle\eta_2|\rho|\eta_1\rangle
|\eta_1-\eta_2|^2\;\;.\label{eq35}\end{equation}
This gives a measure of the mean squared variance of the distribution
$|\rho(\eta_1,\eta_2)|^2$ along the off-diagonal. Because this measure
has dimensions of $|\eta|^2$, comparisons of $\hat{\Omega}$ 
between bases are made complicated through the different
length scales associated with each basis. For a comparison between
bases one should use the dimensionless measure (\ref{eq27}). We also
note that while the discrete version of (\ref{eq35}) vanishes for
$\rho\rightarrow$diagonal (eg. spin 1/2 particle), for
\begin{eqnarray}
\rho &=&|\psi\rangle\langle\psi|\;\;,\nonumber\\
2|\psi\rangle &=&|\alpha\rangle+|-\alpha\rangle\;\;,\label{state}
\end{eqnarray}
the continuous version
of (\ref{eq35}) evaluated in the coherent state basis increases like
$|\alpha|^2$. 
Although (\ref{eq27}) and (\ref{eq35}) seem to be natural
candidates for a measure  of ``diagonality'' in the EIS sense we see
that the prototypical state (\ref{state}) is not diagonal nor even
approximately diagonal in the coherent state basis. It may be said
that since the ripples in the Wigner function decrease in size as
$|\alpha|^2\rightarrow\infty$ one should treat this state in this
limit as ``interferenceless'' or ``classical'', This I believe is
incorrect. The Wigner function for this state 
always possesses negative regions for all
$|\alpha|\neq\infty$.  
The interference represented by these would be manifested
in the expectation values of observables which have support in these
regions. The effects of EIS are to cause these negative regions to
disappear in a finite time. However, as pointed out by Bell,
positivity of the Wigner function is a necessary but {\em not}
sufficient condition that the Wigner distribution correctly represents a true
probability distribution \cite{BELL:SPEAKABLE}. This connection will
be made clearer in \cite{TWAMLEY:1993d}. Finally we can use the
relation
\begin{equation}
\langle x|\rho| y\rangle=\int\,dp\,e^{ip(y-x)/\hbar}
W(p,\frac{x+y}{2})\;\;,\label{eq30}\end{equation}
to rewrite (\ref{eq35}) as

\widetext\begin{eqnarray}
&&\hat{\Omega}_{\eta}=\frac{1}{\int\,dpdx\,W^2(p,x)}\times\nonumber\\
&&\left\{
1+8\hbar\omega\int\,dpdx\left[
W(p,x)\left( \frac{d^2}{\omega^2dx^2}+\frac{d^2}{dp^2}\right)W(p,x)-
\left(\frac{d}{\omega
dx}W\right)^2-\left(\frac{d}{dp}W\right)^2\right]
\right\}\;\;,\label{eq30a}\end{eqnarray}\narrowtext

where we have used
\begin{equation}
{\rm
Tr}\,\rho^2=\int\,\frac{d^2\alpha}{\pi}\,W^{2}(\alpha,\alpha^{*})=
\int\,\frac{dpdx}{2\pi\hbar}W^{2}(p,x)\;\;.\end{equation}

In what follows we will use $\Omega_{\eta}$, (\ref{eq27}) as a measure of
diagonality. Taking $\rho(t=0)=|\alpha_0\rangle\langle\alpha_0|$ we can
compute
\begin{equation}
\Omega_{x}(t)=\Omega_{p}(t)=\Omega_{\alpha}^2=
\frac{\sigma_{-}^2}{\sigma_{+}^2}=\frac{1}{(2\kappa(t)-1)^2}
\;\;.\label{EIS}\end{equation}
Recalling that $\kappa(t)=1+\bar{n}(1-e^{-\gamma t})$, (\ref{eq14}), 
this result 
would indicate that we obtain a greater degree of diagonalisation
in the $x$ and $p$ basis than in the coherent state basis $\alpha$. At
$t=0$, since the system begins in a pure state, $\Omega=1$ for all
bases. Thus Environmental Induced Superselection has occurred in all
three bases. From (\ref{eq24a}), $|\langle n+m|\rho(t)|n-m\rangle|\sim
e^{-m\gamma t}$ and thus $\rho$ also diagonalises in the number basis,
however the exact form of $\Omega_{n}$ is quite complicated and will 
not be given here. For the
case of a completely general initial state we only quote the result in
the position basis. For
$\rho(t=0)=\sum\,N_{ij}|\alpha_{j}\rangle\langle\alpha_{i}|$ one finds
\widetext\begin{eqnarray}
&&\Omega_{x}=\nonumber\\
&&\frac{\sigma_{-}^2}{\sigma_+^2}\left\{\frac{
\sum_{ijkl}\,N_{ij}N_{kl}\langle\alpha_i|\alpha_j\rangle
\langle\alpha_k|\alpha_l\rangle e^{A_+^2\sigma_+^2+A_-^2\sigma_-^2+B}
\left[1+2A_-^2\sigma_-^2\right]}
{\sum_{ijkl}\,N_{ij}N_{kl}\langle\alpha_i|\alpha_j\rangle
\langle\alpha_k|\alpha_l\rangle e^{A_+^2\sigma_+^2+A_-^2\sigma_-^2+B}
\left[1+2A_+^2\sigma_+^2+8\langle x\rangle(\frac{\langle
x\rangle}{\sigma_+^2}-A_+)\right]}\right\}\;\;,\label{19a}\end{eqnarray}
\narrowtext
where
\begin{eqnarray}
A_-&=&\frac{-1}{2\sqrt{\sigma_-\sigma_+}}
\left[\alpha_i^*(t)-\alpha_k^*(t)+\alpha_l(t)-\alpha_j(t)\right]\;\;,\\
A_+&=&\frac{\sqrt{\sigma_+\sigma_-}}{2\sigma_+^2}
\left[\alpha_i^*(t)+\alpha_k^*(t)+\alpha_j(t)+\alpha_l(t)\right]\;\;,
\end{eqnarray}
and
\begin{equation}
B=-\left[(\alpha_i^*(t)+\alpha_j(t))^2+(\alpha_k^*(t)+\alpha_l(t))^2\right]
\left(\frac{\sigma_-}{2\sigma_+}\right)\;\;.\end{equation}
A similar result holds for $\Omega_p$ where we replace $A_\pm$ with
\begin{eqnarray}
A_+ &=&
\frac{i}{2\sqrt{\sigma_+\sigma_-}}\left[\alpha_i^*(t)-\alpha_k^*(t)
+\alpha_j(t)-\alpha_l(t)\right]
\;\;,
\nonumber\\
A_- & = & \frac{-i\sqrt{\sigma_+\sigma_+}}{2\sigma_+^2}
\left[\alpha_i^*(t)+\alpha_k^*(t)-\alpha_j(t)-\alpha_l(t)\right]\;\;.
\end{eqnarray}

To obtain a clearer understanding of $\Omega$ let us compute
$\Omega_x$ for the particular state 
$\rho_2=
\frac{1}{2}|\alpha_0\rangle\langle\alpha_0|+\frac{1}{2}|-\alpha_0\rangle
\langle -\alpha_0|$ where
$\alpha_0=(\omega\bar{x}+i\bar{p})/\sqrt{2\hbar\omega}$. Denoting
$\rho_1=|\alpha_0\rangle\langle\alpha_0|$ we find that
$\Omega_{2\,x}=\Omega_{1\,x}\hat{\Omega}_{2\,x}$ where
$\Omega_{1\,x}=\sigma_{-}^2/\sigma_{+}^2$, (from above) and
\begin{equation}
\hat{\Omega}_{2\,x}=\left[
\frac{1+\frac{1}{2}(1-\frac{2\bar{p}^2\sigma_-^2}{\hbar^2})
e^{-4\bar{x}^2/\sigma_+^2-\bar{p}^2\sigma_-^2/\hbar^2}}
{1+\frac{8\bar{x}^2}{\sigma_+^2}+
e^{-4\bar{x}^2/\sigma_+^2-\bar{p}^2\sigma_-^2/\hbar^2}}\right]\;\;,
\label{omegax}\end{equation}
where $\sigma_\pm$ are defined in (\ref{xy1}) and evaluated at $t=0$.
Since $\hat{\Omega}_{2\,x}\le 1$ we see that the coherence of a
mixture of two Glauber states is less than the coherence of just one.
The \mbox{$x$-coherence} of the mixture $\rho_2$ decreases inversely as
$(\bar{x}/\sigma_{+})^2$. Thus packets which are separated in
position by amounts significantly greater than their half widths
possess lower $x$-coherence. Also $\hat{\Omega}_{2\,x}$ depends very
weakly on the momentum $\bar{p}$ except for small $\bar{x}$ where the
overcompleteness property of the Glauber states cause
interference-like behaviour in the $x$ basis. A plot of (\ref{omegax})
is given in Fig 1. 

The dynamical evolution of $\Omega_{2\,x}$ is
given again by equation (\ref{omegax}) with $\sigma_{\pm}(t)$ given
by (\ref{xy1}) while
\begin{eqnarray}
\bar{x}(t)&=&e^{-\gamma t/2}\left[\bar{x}(0)\cos \omega
t+\frac{\bar{p}(0)}{\omega} \sin \omega t\right]\;\;,\\
\bar{p}(t)&=&e^{-\gamma t/2}\left[\bar{p}(0)\cos\omega t-\omega
\bar{x}(0)\sin\omega t\right]
\;\;.\\
\end{eqnarray}
Essentially, the evolution begins at $t=0$ 
with the two Glauber states located symmetrically about the origin at
positions $\pm\alpha_0\in\Bbb{C}$. Since
the Glauber states are overcomplete there is some overlap between the
two. As time progresses, the state, represented loosely as two
superposed gaussians in the Wigner distribution, spirals in towards
the origin while the half widths of each gaussian spreads with time
for nonzero $\bar{n}$. The system motion complicates the
interpretation and instead, one can return to the interaction picture.
In the interaction picture, for a fixed choice of $\alpha_0(t=0)$ where
$\bar{x}\gg\sqrt{2\hbar/\omega}$, 
$\Omega_{2\,x}$ decreases below the value $\Omega_{1\,x}$ due to the
large separation in position between the two gaussians. However, as
$t\rightarrow\infty$, $\hat{\Omega}_{2\,x}\rightarrow 1$ and thus the
coherence of $\rho_2(t)$ tends to the same value as the coherence of
$\rho_1(t)$ in this limit. This is not surprising as the final states
are the same. As the bath temperature is raised the time for
$\Omega_{2\,x}$  to reach a given value decreases since the
gaussian's half widths become larger and thus $\Omega_{1\,x}$
decreases faster. Also, we note that when the $\bar{x}$ separation is
large, the state may possess a {\em lower} coherence than in the
equilibrium state at $t=\infty$. This phenomena is also seen is the
relaxation behaviour of squeezed states in a thermal bath. 
The coherence behaviour of extended quantum states
and the relation of the $\Omega$ measure of coherence to more
information theoretic measures will be treated in another paper
\cite{TWAMLEY:1993d}.
\begin{center}
\graphics{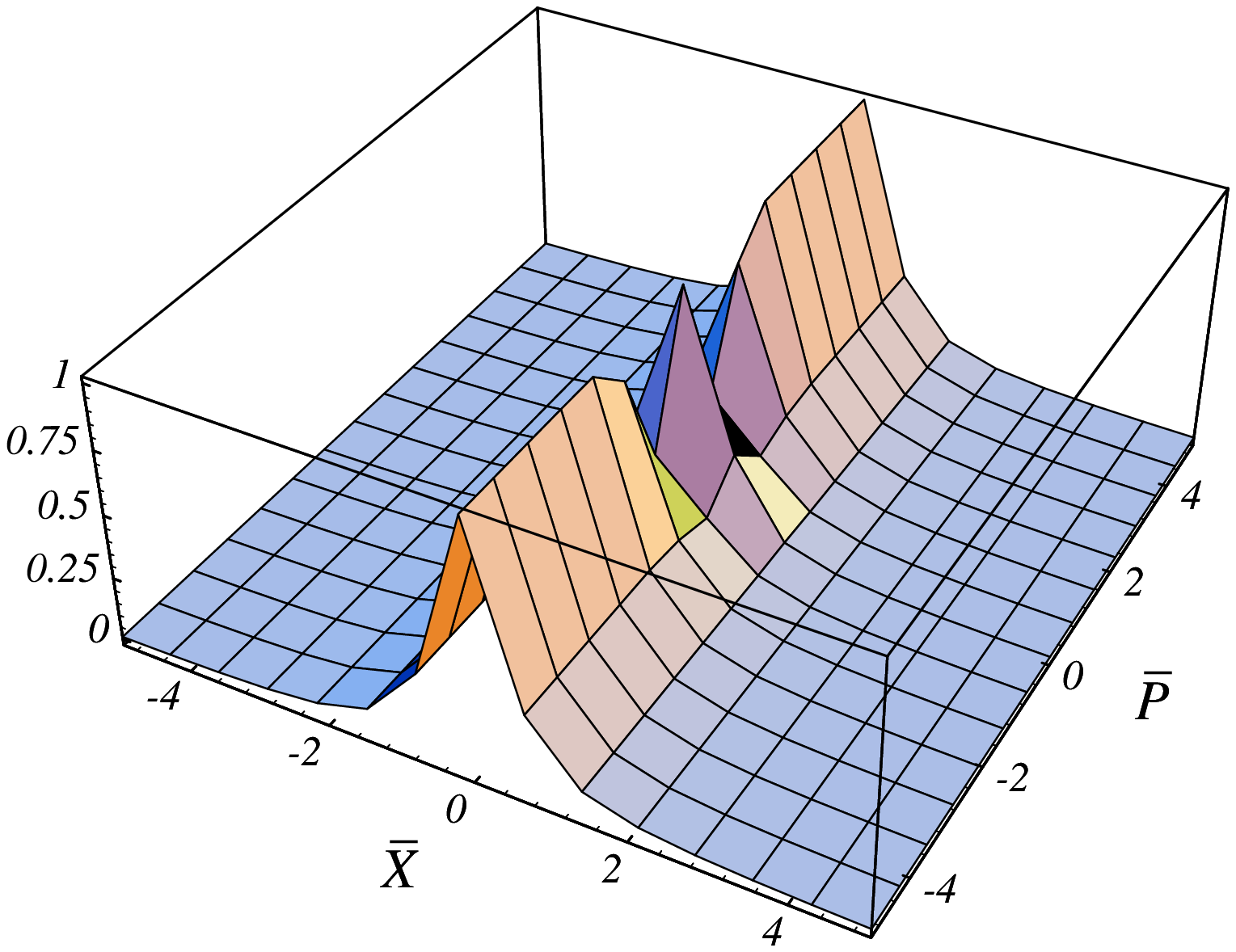}{5}{7}{-2}{-10}{.5}
\end{center}

\begin{center}
{\parbox{15cm}{\small\center  Fig 1.
Surface  plot of $\Omega_x$ for state $\rho_2$ with $\omega=1$ and $\hbar=1$.
}}
\end{center}
To conclude and summarize: we have described the dynamics of the
reduced density matrix of a harmonic oscillator linearly coupled to an
infinite bath of oscillators in the quantum optical regime. We have
expressed the results in the position, momentum, number and coherent
state basis. After considering a number of proposals for a measure of
the ``diagonality'' or degree of coherence of the state $\rho$ we have
advanced $\Omega_\eta$ (\ref{eq27}) as a natural measure. We
calculated this quantity in the above mentioned bases for an initial
Glauber state and found that coherence was lost in each basis. The coherence
loss rate for this initial state was found to be identical in the $x$
and $p$ basis  while the loss rate in the coherent state basis was
smaller. Expressions for the coherence of a general initial state were
found in both the $x$ and $p$ bases. Finally, the position coherence for
an initial mixture of two coherent states was calculated and was found
to be more ``diagonal'' than that of a single coherent state,
especially of the position separation between the two coherent states is large.
Although the measure $\Omega_\eta$ may seem to be a ``natural''
generalization of the early definitions \cite{MORIKAWA:1990} it would be
interesting to understand the information theoretic aspects of such
measures. Of more interest would be the link between the
``diagonality'' of $\rho$ and the representation of $\rho$ by a true
probability distribution. This work is in progress \cite{TWAMLEY:1993d}.

\section{Consistent Phase-Space Histories}
Consistent Histories is an alternate paradigm developed by Hartle 
and Gell-Mann \cite{HARTLE:1993a,HARTLE:1989,OMNES:1992,GELL-MANN:1992d}
 which attempts to describe the
emergence of the ``Quasiclassical'' world from the quantum world. The
primary ingredients of this paradigm are coarse grained histories and
the decoherence functional. Essentially, a coarse grained history is a
partial description of the temporal evolution of the total
system+environment where one has ``coarse grained'' away information
either by (i) ignoring some of the coordinates for all time (ie. the
environment) (ii) only describing  the configuration of the priveliged
coordinates at certain times (iii) projecting  the 
privileged system coordinates at specified time on to quantities of
interest eg. binning, averaging etc. The resulting coarse grained
histories may be described in some cases (those treated here) 
by history independent
temporally ordered strings of projectors $P^\alpha_i$, where the
$P_i^\alpha$ are a complete and exclusive set of projectors on to a set
of alternatives labeled by $\{\alpha\}$ and where $P_i^\alpha$ is the
ith projector in the set $P^{\{\alpha\}}$. In the past, calculations
have concentrated on sequences on projectors in position using the
Calderia-Legget model \cite{DOWKER:1992,PAZ:1992a}. In this section we
expand these calculations to projectors on phase-space and consider
coarse grained phase-space 
histories of the privileged system harmonic oscillator
in the quantum optical regime. In the Consistent Histories paradigm a
probability is assigned to a coarse grained history via the rule 
\begin{equation}
P(h)\equiv {\rm Tr}[P_{h}\rho P_h^\dagger ]\;\;,\label{cheq1}
\end{equation}
where $P_h$ is the temporally ordered string of projectors
corresponding to the coarse grained history $h$ ie.
\begin{equation}
P_h\equiv 
P_{i_N}^{\alpha_N}{\cal U}(\Delta t_N)
P_{i_{N-1}}^{\alpha_{N-1}}
{\cal U}(\Delta t_{N-1})\cdots P_{i_1}^{\alpha_1}
\;\;,\label{cheq2}
\end{equation}
and where $\cal U$ is the propagator for the full system-environment.
We now assume that these projectors factor out the environment and can thus be
written in the form $P_i^\alpha=\tilde{P}_{i\,S}^\alpha\otimes I_{E}$.
In the Markov regime the dynamics of system's reduced density operator 
is purely local in time and $\rho_{ S}\equiv {\rm Tr}_{ E}\,\rho$
 satisfies 
partial differential equation which is first order in time and contains
only system quantities.
In this regime
we can take the trace in (\ref{cheq1}) over
then environment and arrive at
\begin{equation}
P(h)={\rm Tr}_S\left[\tilde{P}^{\alpha_{N}}_{i_{N}}\,
e^{\int_{t_{N-1}}^{t_{N}}\,{\cal
L}dt}\left[\tilde{P}_{i_{N-1}}^{\alpha_{N-1}}\cdots\rho_{S}\cdots\right]\right]
\;\;,\end{equation}
where $\tilde{P}$ acts only on the system's Hilbert space,
$\rho_{S}$ is the reduced density operator for the system and we have
again used (\ref{eq1.1}) to evolve the system's reduced density operator
 from $t_{i-1}$ to $t_i$. A similar method is used in \cite{PAZ:1993}.
Thus in the Markov regime the effects of the coupling to the
environment are taken care of through the superoperator $\cal L$ and
the calculation (\ref{cheq1}) reduces to a calculation on the system
only.

At the heart of this formalism is the requirement that these
probabilities satisfy the sum rule
\begin{equation}
P(h)=\sum_{\tilde{h}\in h}\,P(\tilde{h})\;\;,\label{cheq3}
\end{equation}
where $\tilde{h}$ is a finer graining of $h$. For the sum rule
(\ref{cheq3}) to hold with the graining $\tilde{h}$ we must have
\begin{equation}
{\rm Re}\,{\rm D}[\tilde{h}^\prime,\tilde{h}]=0\qquad\forall\;
\tilde{h}^\prime,\;\tilde{h}\in h\;\;,\tilde{h}^\prime\neq\tilde{h}\;\;,
\label{cheq4}\end{equation}
where
\begin{equation}
{\rm D}[\tilde{h}^\prime,\tilde{h}]={\rm Tr}[P_{\tilde{h}}\rho
P_{\tilde{h}^\prime}^\dagger ]\;\;,\label{decoherence_functional}
\end{equation}
is known as the Decoherence Functional \cite{HARTLE:1989}.
Condition (\ref{decoherence_functional}) is often termed {\em Weak
Consistency}. A stronger, although sometimes argued more physical,
condition is 
\begin{equation}
{\rm D}[\tilde{h}^\prime,\tilde{h}]=0\qquad\forall\;
\tilde{h}^\prime,\;\tilde{h}\in h\;\;,\tilde{h}^\prime\neq\tilde{h}\;\;,
\end{equation}
and is called {\em Medium Consistency}.
A major goal in this
approach is the identification of all those sets of histories which
are exactly consistent and from this, an identification of those
histories which yield a ``Quasiclassical'' world. We do not wish to
give an in-depth review of this formalism and refer the reader the
excellent articles of Gell-Mann and Hartle 
\cite{HARTLE:1993a,HARTLE:1989,OMNES:1992,GELL-MANN:1992d}.

It has been recognized that sets of histories which are {\em exactly}
consistent may be characterized by very exotic and un-physical system
variables \cite{ALBRECHT:1992}. To examine the consistency properties of
histories parametrised by (perhaps) more interesting and physical
variables (ie. position and momentum) a measure of the degree of
consistency of a set of histories $\{\tilde{h}\}$ has been advanced by
Dowker and Halliwell \cite{DOWKER:1992}
\renewcommand{\arraystretch}{.5}
\begin{equation}
\epsilon\equiv
\begin{array}{c}
{\rm max} \\
\tilde{h}^\prime,\; \tilde{h}\in\{h\}\\
\tilde{h}^\prime\neq\tilde{h}
\end{array}\frac{{\rm Re}{\rm D}
[\tilde{h}^\prime,\tilde{h}]}
{\sqrt{{\rm D}[\tilde{h},\tilde{h}]{\rm D}
[\tilde{h}^\prime,\tilde{h}^\prime]}}\;\;.\label{epsilon}\end{equation}
As $\{\tilde{h}\}$ becomes more consistent $\epsilon$ decreases.
Dowker and Halliwell have also argued that in general $\epsilon$
decreases as the graining size is increased. Only for very artificial
grainings, where the phases of the constituent histories in
the coarser graining add constructively, does $\epsilon$ remain
unchanged or actually increase. This point will be important later.
As mentioned above, all previous calculations have looked at histories on
position space. However we normally consider classical dynamics as
operating within phase-space. It would be of great interest to see if
phase-space histories of the system become more consistent with the
addition of an interaction with an external bath. In the previous
section on Environmental Induced Superselection we saw that the
quantum optical model gave EIS in both the $x$ and $p$ bases. Here we wish
to see whether phase-space histories for the privileged harmonic
oscillator displays consistency and if so, to what degree?  How does
the degree of consistency
depend on the model parameters ie. $\gamma$, $\bar{n}$ graining
etc? In the following we will consider two-time projected histories on
the phase-space of the privileged oscillator. We will again couple the
system to an external bath and use the solutions of the master
equation (\ref{eq1.1}) found in section II to evolve $\rho_{S}$ between
projections. We will calculate the Decoherence Functional and the
degree $\epsilon$ of consistency of these histories. This $\epsilon$
will be a function of the equilibrium number of photons per mode in the
bath ($\bar{n}=1/(\exp\frac{\hbar\omega}{kT}-1)$), the time $t$, the
initial state of the system $\rho(t=0)$ and the phase-space projectors
chosen (ie. the graining). We will choose gaussian type ``projectors''
on phase-space and will find that $\epsilon$ decreases with time $t$,
increasing bath temperature $T$ and final grain size $\sigma_2$ but
anomalously decreases with initial grain size $\sigma_1$. This
anomalous behaviour can also be seen in other calculations of
Decoherence Functionals for similar models
\cite{DOWKER:1992,PAZ:1992a}. We conclude that the introduction of the
bath causes the phase-space histories to become more consistent and
less peaked about the classical path. Finally we note that Consistent
Histories calculations tend to give rather lengthy and hard to
interpret results. We have found that working with this model has
resulted in a more transparent description.

\subsection*{Two-Time Phase Space Histories}
We wish to calculate the decoherence functional for two-time projected
histories on phase-space, that is
\begin{eqnarray}
{\rm D}[\tilde{h},\tilde{h}^\prime] &=&
{\rm D}[\alpha_3,\alpha_2,\alpha_1,\sigma_2,\sigma_1,t]\nonumber\\
& =&
{\rm Tr}\left[P(\alpha_3,\sigma_2)\,e^{\int_{t_1}^{t_2}\,{\cal L}dt}
\left[P(\alpha_2,\sigma_1)\rho(t_1)P^\dagger(\alpha_1,\sigma_1)\right]\right]
\;\;,\label{cheq6}\end{eqnarray}
where the coarse grained history is defined by a projector
$P(\alpha,\sigma)$ on the phase-space of the system oscillator at
$\alpha\in\Bbb{C}$ 
of ``width'' $\sigma$. The trace in
(\ref{cheq6}) is over the system's Hilbert space. We will assume, 
as in \cite{DOWKER:1992,PAZ:1992a}, that $P$ is a {\em
quasi-projector}. 
In particular we will take $P$ to be  gaussian  with
\begin{equation}
P(\alpha,\sigma)=\frac{1}{\sigma^2}\int\,\frac{d^2\beta}{\pi}
e^{-\frac{|\beta-\alpha|^2}{\sigma^2}}|\beta\rangle\langle\beta|\;\;,
\label{cheq7}
\end{equation}
where $|\beta\rangle$ is a Glauber coherent state. These projectors are
complete ie. $\int\,d^2\alpha/\pi\,P(\alpha,\sigma)=1$, but are not
exclusive
\begin{equation}
P(\alpha_1,\sigma)P(\alpha_2,\sigma)\neq
\delta(\alpha_1-\alpha_2)P(\alpha_2,\sigma)\;\;.\label{not_projectors}
\end{equation}
Using such gaussian ``quasi-projectors'' makes the calculation analytically
tractable. In fact, since true projectors on phase-space do not exist
\cite{OMNES:1989b} the best one can do is to use such complete but
nonexclusive ``quasi-projectors''. Indeed, these types of
``quasi-projections'' play a significant role in the more rigorous ``Effects
 and Operations'' theory of quantum measurement
\cite{CAVES:1986}. To create other quasi-projectors which are more
localized in phase-space one could use instead the marginally
overcomplete set of coherent states $\{|\alpha\rangle\}_{VN}$ where
one places an ordinary Glauber coherent state $|\alpha\rangle$ at each
point of a \mbox{von Neumann} lattice in phase space
\cite{PERELOMOV:1971,BARGMANN:1971}.

We shall calculate (\ref{cheq6}) for an arbitrary $\rho(t_1)$ by
decomposing $\rho(t_1)$ into coherent states and using (\ref{eq20})
to evolve to $t_2$. We will then consider a $\rho(t_1)$ resulting from
the propagation from a time $t_0<t_1$ of $\rho(t_0)$ which we shall
choose to be $\rho(t_0)=|\alpha_0\rangle\langle\alpha_0|$. We will
obtain the decoherence functional ${\rm D}$ and from this
calculate the degree of consistency $\epsilon$ and peaking $\cal P$
about the classical trajectory. We will also calculate $\Omega_h$, a
measure similar to (\ref{eq27}), which measures the coherence in 
${\rm D}$ when considered as a density matrix. We find that
$\epsilon$ and $\Omega_h$ behave similarly with respect to their
dependence on the parameters in the model.

To begin, we take
$\rho(t_1)=\sum_{ij}N_{ij}|\alpha_j\rangle\langle\alpha_i|$ where we
have normalized according to
$\sum_{ij}N_{ij}\langle\alpha_j|\alpha_i\rangle=1$. Using (\ref{appendix1}),
\begin{equation}
\int\,\frac{d^2\alpha}{\pi} e^{-A|\alpha|^2+B\alpha^*+C\alpha}=
\frac{1}{A}e^{\frac{BC}{A}}\;\;,\label{cheq9}\end{equation}
and the definition of a coherent state $|\alpha\rangle=e^{\alpha
a^\dagger -\alpha^* a}|0\rangle$ we can compute
\begin{equation}
P(\alpha_2,\sigma_1)\rho(t_1)P^\dagger (\alpha_1,\sigma_1)=
\frac{1}{s_1^2}\sum_{ij}{\bbox{\Bbb F}}
(\alpha_1,\alpha_2,\alpha_i,\alpha_j,\sigma_1)|\zeta_j\rangle\langle\eta_i|
\;\;,\label{cheq10}\end{equation}
where 
\widetext
\begin{equation}
s_1=1+\sigma_1^2\;,\qquad\qquad
s_1\zeta_j=\sigma_1^2\alpha_j+\alpha_2\;,\qquad\qquad 
s_1\eta_i=\sigma_1^2\alpha_i+\alpha_1\;\;,\label{cheq11}\end{equation}
and
\begin{eqnarray}
{\bbox{\Bbb F}}(\alpha_1,\alpha_2,\alpha_i,\alpha_j,\sigma_1)=
\exp\left\{-\frac{1}{2s_1^2}
\right.&& 
\left[ (|\alpha_i|^2+|\alpha_j|^2+|\alpha_1|^2+|\alpha_2|^2)
(1+2\sigma_1^2)\right.\nonumber\\
&&\left.\left.+(\alpha_2^*\alpha_j+\alpha_1\alpha_i^*)(2+3\sigma_1^2)
+\sigma_1^2(\alpha_2\alpha_j^*+\alpha_1^*\alpha_j)\right]\right\}\;\;.
\label{cheq12}\end{eqnarray}
Using (\ref{eq20}) we can evolve (\ref{cheq10}) from $t_1$ to 
$t_2=t_1+\Delta t_2$ and denoting this by $\rho_{eff}(t_2)$ we get
\begin{equation}
\rho_{eff}(t_2)=\int\,\frac{d^2\alpha  d^2\beta}{\pi^2}\,
\langle\alpha|\rho_{eff}(t_2)|\beta\rangle\,|\alpha\rangle\langle\beta|\;\;,
\label{cheq13}\end{equation}
where
\begin{eqnarray}
&&\langle\alpha|\rho_{eff}(t_2)|\beta\rangle=\nonumber\\
&&\frac{1}{s_1^2}\sum_{ij}
\frac{{\bbox{\Bbb F}}(\alpha_1,\alpha_2,\alpha_i,\alpha_j,\sigma_1)}
{\kappa(\Delta t_2)}
\exp\,\left\{
-\frac{1}{\kappa(\Delta t_2)}\left(\alpha-\eta_i(\Delta t_2)\right)^*
\left(\beta-\zeta_j(\Delta t_2)\right)\right\}
\langle\alpha|\beta\rangle\langle\eta_i|\zeta_j\rangle\;\;,\label{cheq14}
\end{eqnarray}
where $\zeta_j(t)\equiv e^{-(\gamma/2+i\omega)t}\zeta_j$ and
$\eta_i(t)\equiv e^{-(\gamma+i\omega)t}\eta_i$.
The final decoherence functional is thus
\begin{equation}
{\rm D}(\alpha_3,\alpha_2,\alpha_1,\sigma_2,\sigma_1,t_2)=
\frac{1}{\sigma_2^2}\int\frac{d^2\gamma}{\pi}e^{-\frac{|\alpha_3-\gamma|^2}
{\sigma_2^2}}\,\langle\gamma|\rho_{eff}(t_2)|\gamma\rangle\;\;,\label{cheq15}
\end{equation}
and is thus a gaussian smearing of the Q function of
$\rho_{eff}(t_2)$. Note that although Q functions are strictly positive for
true density matrices, when $\alpha_1\neq\alpha_2$, $\rho_{eff}(t_2)$
is {\em not} a true density matrix and thus ${\rm D}
[\alpha_3,\alpha_2,\alpha_1\neq\alpha_2]$ may be complex. To
calculate (\ref{cheq15}) is quite tedious but one finally arrives at
\begin{eqnarray}
{\rm D} = && \sum_{ij}\frac{N_{ij}}{\sigma_1^2}
{\bbox{\Bbb F}}(\alpha_2,\alpha_1,\alpha_i,\alpha_j,\sigma_1)
\frac{\langle\eta_i|\zeta_j\rangle}{s_2}\times\nonumber\\
&&\exp\left\{
-\frac{|\alpha_3|^2}{\sigma_2}
-\frac{\eta_i^*(\Delta t_2)\zeta_j(\Delta t_2)}{\kappa_2}
+\frac{(\kappa_2\alpha_3+\sigma_2^2\zeta_j(\Delta t_2))
(\kappa_2\alpha_3+\sigma_2^2\eta_i(\Delta t_2))^*}
{s_2\sigma_2^2\kappa_2 }\right\}\;\;,\label{cheq16}\end{eqnarray}
where $\kappa_2\equiv\kappa(\Delta t_2)$ and 
$s_2\equiv\kappa(\Delta t_2)+\sigma_2^2$. For simplicity we will now
turn to the interaction picture and consider our projectors to be
stationary in this picture ie. $\alpha_1\rightarrow\alpha_1
e^{-i\omega t}$. With this, all $\omega$ dependence drops out of
(\ref{cheq16}). With a bit of rearranging (\ref{cheq16}) can be
written in the more convenient form
\widetext
\begin{eqnarray}
{\rm D} =
\sum_{ij}\frac{N_{ij}}{s_1^2s_2}\exp &&\left\{
-\frac{|\alpha_i|^2}{2}-\frac{|\alpha_j|^2}{2}
+\frac{(s_1-1)^2}{s_1^2}\alpha_i^*\alpha_jG
+\alpha_i^*{\bbox{\Bbb E}}(\alpha_1,\alpha_2,\alpha_3)
+\alpha_j{\bbox{\Bbb E}}^*(\alpha_2,\alpha_1,\alpha_3)\right.\nonumber\\
&&\left. 
-\frac{|\alpha_3|^2}{s_2}-\frac{|\alpha_1|^2}{s_1}-\frac{|\alpha_2|^2}{s_1}
+\frac{\alpha_1^*\alpha_2}{s_1^2}G
+(\alpha_3^*\alpha_2+\alpha_3\alpha_1^*)\frac{e^{-\gamma \Delta
t/2}}{s_1s_2}
\right\}\;\;,\label{cheq17}\end{eqnarray}\narrowtext
where $G\equiv 1-e^{-\gamma \Delta t_2}/s_2$ and
\begin{equation}
{\bbox{\Bbb E}}(\theta_1,\theta_2,\theta_3)\equiv
\frac{\theta_1}{s_1}+\frac{\sigma_1^2\theta_2G}{s_1^2}
+\frac{\sigma_1^2\theta_3e^{-\gamma\Delta t_2/2}}{s_1s_2}\;\;.
\label{cheq19}\end{equation}\narrowtext

We are now ready to choose
$\rho(t_0)=|\alpha_0\rangle\langle\alpha_0|$. From (\ref{eq1.1}) 
we have
\begin{equation}
\rho(t_1)=\int\,\frac{d^2\alpha_j d^2\alpha_i}{\pi^2}\,
\langle\alpha_j|\,e^{\int_{t_0}^{t_1}\,{\cal
L}dt}\rho(t_0)|\alpha_i\rangle\,
|\alpha_j\rangle\langle\alpha_i|\;\;,\end{equation}
and using (\ref{eq20}) yields
\widetext\begin{equation}
\rho(t_1)=\int\,\frac{d^2\alpha_jd^2\alpha_i}{\kappa(\Delta t_1)\pi^2}\,
e^{-\frac{1}{\kappa(\Delta t_1)}
(\alpha_j-\alpha_0e^{-\gamma \Delta t_1/2})^*
(\alpha_i-\alpha_0r^{-\gamma \Delta t_1/2})}
\langle\alpha_j|\alpha_i\rangle\,|\alpha_j\rangle\langle\alpha_i|\;\;,
\label{cheq20}\end{equation}
where $\Delta t_1=t_1-t_0$.
Since the action of the decoherence functional is linear we can use 
(\ref{cheq17}) to obtain the action on 
$|\alpha_j\rangle\langle\alpha_i|$ and  can then 
perform the integrations in (\ref{cheq20}) to
compute the complete decoherence functional. We finally get
\begin{equation}
{\rm D}=\frac{1}{s_2\Upsilon}\exp\{ -A|\Delta_1|^2
-B|\Delta_2|^2 +C(\Delta_1^*\Delta_2+\Delta_1\Delta_2^*) -D|z|^2-i{\rm
Im}{\bbox{\Bbb A}}\}\;\;,\label{cheq21}\end{equation} where the
definitions of the various quantities appearing in (\ref{cheq21}) are
\widetext
\begin{equation}
\begin{tabular}{ll}
$\Upsilon=s_1^2\kappa_1-\Phi (s_1-1)^2G\;,\qquad\qquad$&
$4\Upsilon A=2s_1-1+\frac{1}{s_2}e^{-\gamma \Delta t_2}(1-\Phi)\;,$\\
$\Phi=\bar{n}(1-e^{-\gamma \Delta t_1})\;,$ & \\
$ \Delta_1^*=\alpha_1+\alpha_2-2\alpha_0e^{-\gamma \Delta t_1/2}\;,$ &
$s_2\Upsilon B=s_1^2-\Phi\;,$\\
$\Delta_2^*=\alpha_3-\alpha_0e^{-\gamma(\Delta t_2+\Delta t_1)/2}\;,$ &
\\
$ z=\alpha_1-\alpha_2 \;,$& $2s_2\Upsilon C=(s_1-\Phi)
e^{-\gamma\Delta t_2/2}\;,$\\
$G=1-\frac{e^{\gamma \Delta t_2}}{s_2}\;,$ & \\
$ s_1=1+\sigma_1^2\;,$ &
$4\Upsilon D=1+2s_1-\frac{1}{s_2}e^{-\gamma \Delta t_2}
+\Phi\left[4-e^{-\gamma \Delta t_2}\frac{(3-2s_1)}{s_2}\right]\;,$\\
$ s_2=\sigma_2^2+\kappa_2 \;,$ & \\
$ \kappa_2=1+\bar{n}(1-e^{-\gamma \Delta t_2}) \;,$ &
$ 2\Upsilon{\bbox{\Bbb A}}=z\left\{
\left[ 1-\frac{1}{s_2}e^{-\gamma\Delta t_2}(1+\Phi)\right]\Delta_1^*
+\frac{2}{s_2}[s_1+\Phi]e^{-\gamma\Delta t_2/2}\Delta_2^*\right\}\;,$\\
$\kappa_1=1+\bar{n}(1-e^{-\gamma \Delta t_1})\;.$ & \\
\end{tabular}\label{cheq22}\end{equation}
\narrowtext

Thus $\Delta_1$ and $\Delta_2$ are the deviations of the projector
positions at times $t_1$ and $t_2$ from the classical path of the
damped coherent state $|\alpha_0\rangle$. The diagonal elements of
${\rm D}$, ie. $z=0$, are peaked along this classical path
($\Delta_1=\Delta_2=0$). We also note that because of the
nonexclusivity of these quasi-projectors the diagonal elements of
the decoherence functional do not sum exactly to one. For a
decoherence functional of the form (\ref{cheq21})  it easy to show
that the degree $\epsilon$ of consistency defined in (\ref{epsilon}) is
\begin{equation}
\epsilon = \max_{\alpha_3\in\Bbb{C}}
\, e^{(C-D)|z|^2}\,\cos\,
{\bbox{\Bbb A}}\;\;.\label{cheq23}\end{equation}
Thus the maximum departure from perfect consistency between a pair of
histories occurs for (i) those histories for which $\alpha_3$ is such
that $|\cos\,{\bbox{\Bbb A}}|=1$ (ii) for pairs of histories
$h,\;h^\prime$ where the quasi-projectors $P(\alpha_1,\sigma_1)$ and
$P(\alpha_2,\sigma_1)$ are separated by their full width at half
maximum (FWHM) 
ie. $z=1+\sigma_1^2=s_1$. From (i), (ii) and (\ref{cheq23}) we get
\begin{eqnarray}
\epsilon&=&\exp\left\{-\frac{
s_1}{2\Upsilon}\left[1-\frac{1}{s_2}e^{-\gamma\Delta
t_2}+\Phi\left(2+e^{-\gamma\Delta
t_2}\frac{(s_1-1)}{s_2}\right)\right]\right\}\nonumber\\
 &\equiv&
\exp\{-\bar{\epsilon}(\sigma_1,\sigma_2,\gamma,\Delta t_2,\Delta
t_1,\bar{n})\}\;\;.\label{cheq24}\end{eqnarray}
For $\bar{\epsilon}\gg 1$ we achieve good consistency. To examine the
degree of peaking about the classical path we follow Dowker and
Halliwell \cite{DOWKER:1992} and 
look at the product of the determinant of the
coefficients of $\Delta_i$  with the FWHM of the initial and final
projectors, that is ${\cal P}\sim (AB-C^2)s_1(1+\sigma_2^2)$. We get
\begin{equation}
{\cal P}=\frac{s_1(1+\sigma_2^2)}{4s_2\Upsilon^2}\left[
(s_1^2-\Phi)(2s_1-1)-\Phi e^{-\gamma\Delta t_2}\frac{(s_1-1)^2}{s_2}
\right]\;\;.\label{cheq25}\end{equation}
For ${\cal P}\gg 1$ we get significant peaking about the classical
path. Equations (\ref{cheq21}, \ref{cheq22}, \ref{cheq24}, \ref{cheq25})
are the main results of this section.

Let us examine the short and long time behaviour of $\bar{\epsilon}$ and
$\cal P$. For $t_2\sim t_1\sim 0$ we have
\begin{equation}
\bar{\epsilon}\approx \frac{1}{2s_1}[1-\frac{1}{s_2}]\;\;.
\label{cheq26}
\end{equation}
Thus as the final graining size $\sigma_2$ increases the degree of
consistency increases. This is in agreement with the general
arguments of Dowker and Halliwell. However, as the initial graining
size $s_1$ is increased, the degree of consistency drops. This
anomalous behaviour is also seen in other calculations
\cite{DOWKER:1992,PAZ:1992a}. In the long time limit, with $\Delta
t_2=\Delta t_1=\tau$ and letting $\tau\rightarrow\infty$ we have
\begin{equation}
\bar{\epsilon}\approx
\frac{(2\bar{n}+1)s_1}{2[s_1^2+\bar{n}(2s_1-1)]}\;\;,\label{cheq27}
\end{equation}
which again decreases with increasing $s_1$. However, the dependence
on bath temperature, through $\bar{n}$, is now apparent and thus the
consistency increases with increasing $T$. In the short time limit the
peaking is
\begin{equation}
{\cal P}\approx\frac{2s_1-1}{4s_1}\;\;,\label{cheq28}\end{equation}
while for long times
\begin{equation}
{\cal P}\approx\frac{s_1(s_1^2-\bar{n})(2s_1-1)}{2[s_1^2+\bar{n}(2s_1-1)]^2}
\left(\frac{1+\sigma_2^2}{1+\bar{n}+\sigma_2^2}\right)\;\;.\label{cheq29}
\end{equation}
The short time and long time limits of $\cal P$ display the expected
behaviour with the peaking getting better with increasing graining size
and decreasing bath temperature and time. Plots of the long time
behaviors of $\bar{\epsilon}$ and $\cal P$ are shown in  Fig 2.
\begin{center}
(a)\hspace{6cm}(b)\\
\graphics{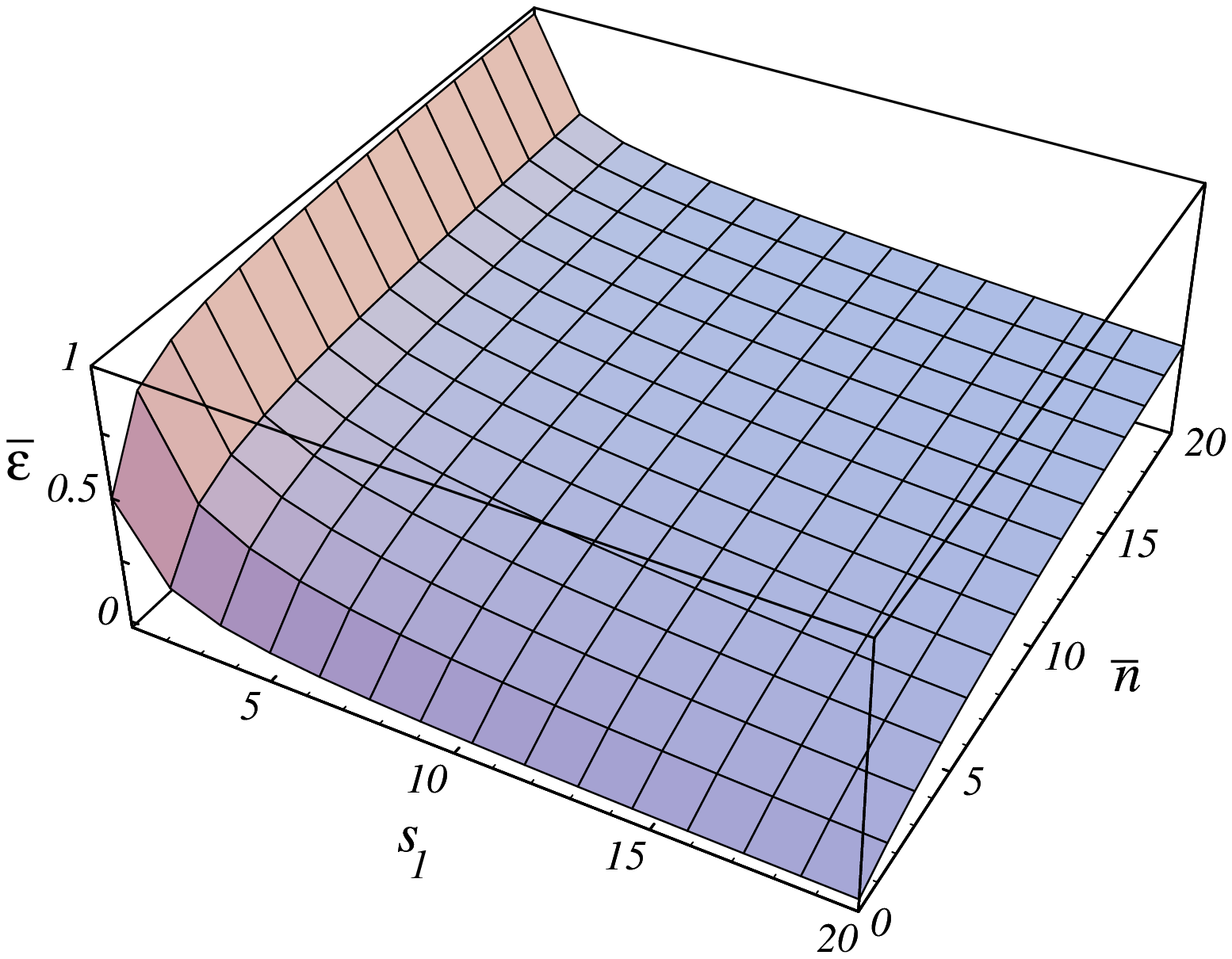}{5}{7}{-2}{-8}{.4}
\graphics{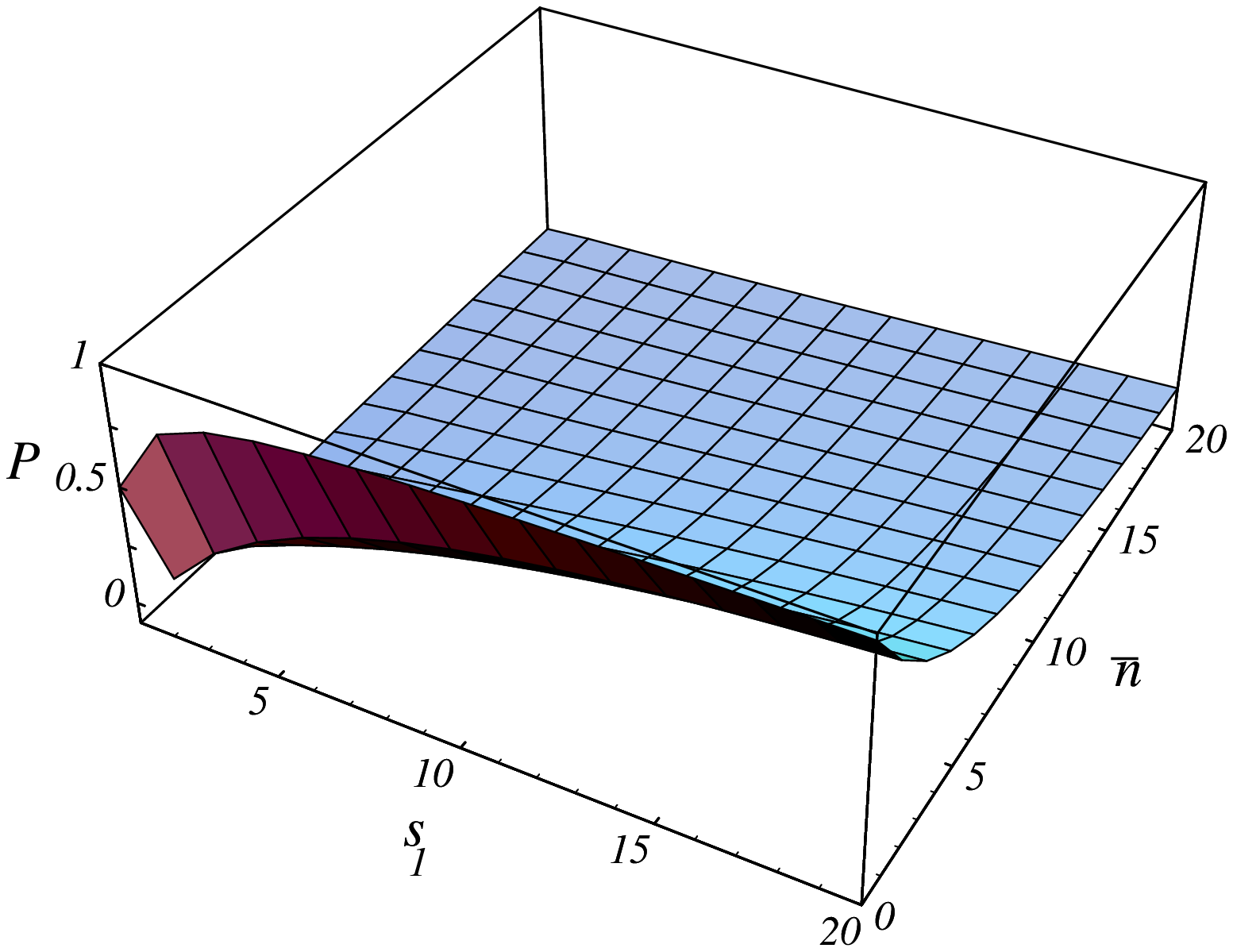}{5}{7}{1}{-8}{.4}
\end{center}
\begin{center}
{\parbox{15cm}{\small  Fig 2.
Dependencies of the 
degree of consistency and degree of peaking in the long time limit 
on the bath temperature
(through $\bar{n}$)  and
initial grain size $s_1$. 
Graph (a) plots (\ref{cheq27}) as a function of $s_1$ and $\bar{n}$,
while graph (b) plots (\ref{cheq29}). We see that increasing the bath
temperature (which increases $\bar{n}$) increases consistency while
decreasing peaking. Increasing the initial grain size $s_1$ {\em
decreases} consistency and increases peaking
}}
\end{center}
We can also examine the degree of consistency by considering the decoherence 
functional as a density matrix in the space of the coarse grained
histories $\{h\}$. With this viewpoint we can adopt the definition
(\ref{eq27}) to give a measure of the concentration of ${\rm
D}[h,h^\prime]$ about the diagonal $h=h^\prime$. For a
decoherence functional of the form
\mediumtext\begin{equation}
{\rm D}\sim N\exp\left[
-A|\Delta_1|^2-B|\Delta_2|^2+C(\Delta_1^*\Delta_2+\Delta_1\Delta_2^*) 
-D|z|^2-i{\rm Im}\,z {\bf\Delta}\right]\;\;,\end{equation}
we can calculate  
\begin{equation}
\Omega_h=\max_{\Delta_2\in\Bbb{C}}\,\frac{
\int\,d^2\Delta_1d^2z\,
{\rm D}[\Delta_2,\Delta_1,z]{\rm D}[\Delta_2,\Delta_1,-z]|z|^2}
{\int\,d^2\Delta_1d^2z\,
{\rm D}[\Delta_2,\Delta_1,z]{\rm D}
[\Delta_2,\Delta_1,-z]|\Delta_1-\langle\Delta_1\rangle|^2}\;\;.\label{cheq30}
\end{equation}\narrowtext
We can motivate this measure by noting that for each value of
$\Delta_2$, ${\rm D}[\Delta_2,\Delta_1,z]$ is very like a
density matrix with the on-diagonals labeled by $\Delta_1$ and the
off-diagonals labeled by $z$. Computing (\ref{cheq30}) for a given
$\Delta_2$ gives a measure of the degree of coherence between all the
possible pairs of histories which have $P(\alpha_3,\sigma_2)$ as their
final quasi-projector. The maximum of (\ref{cheq30}) over $\Delta_2$
gives the largest violation of consistency for the complete set of
histories labeled by both $\Delta_2$ and $\Delta_1$. Before computing
(\ref{cheq30}) we first obtain $\langle \Delta_1\rangle$ for a fixed
$\Delta_2\neq 0$ to be
\begin{eqnarray}
\langle\Delta_1\rangle&=&\int
\frac{d^2\Delta_1}{2\pi}\,N\exp\left[
-A|\Delta_1|^2
-B|\Delta_2|^2
+C(\Delta_1^*\Delta_2+\Delta_1\Delta_2^*)\right]\Delta_1\nonumber\\
 &=&\frac{NC\Delta_2}{A^2}e^ {-(AB-C^2)|\Delta_2|^2/A}\;\;,\end{eqnarray}
which gives $\langle\Delta_1\rangle=0$ when $\Delta_2=0$ as expected.

After some calculation one obtains the maximum violation of
consistency to simply be
\begin{eqnarray}
\Omega_h &=& \frac{A}{D}\;\;,\nonumber\\
 & = & \frac{\left[2s_1-1+\frac{1}{s_2}e^{-\gamma\Delta t_2}(1-\Upsilon)
\right]}{\left[
1+2s_1-\frac{1}{s_2}e^{-\gamma \Delta t_2}+\Upsilon\left(
4-\frac{1}{s_2}[3-2s_1]e^{-\gamma \Delta t_2}\right)\right]
}\;\;.\label{cheq31}
\end{eqnarray}
In the short and long time limits we get
\begin{eqnarray}
\Omega_h(t_1\sim t_2\rightarrow 0) & = &
\frac{2s_1-(1-\frac{1}{s_2})}
{2s_1+(1-\frac{1}{2_s})}\;\leq \;1\;\;,\label{cheq32}\\
\Omega_h(t_1\sim t_2\rightarrow \infty ) &=&
\frac{2s_1-1}{2s_1+1+4\bar{n}}\;\leq\;
\Omega_h(t_1\sim t_2\rightarrow 0)\;\leq\; 1\;\;.
\label{cheq33}\end{eqnarray}
 From (\ref{cheq33}) we see that even when $\bar{n}=0$ the coherence
decays. This is due to the vacuum fluctuations. It is interesting to
note that all dependence on $s_2$ drops out of these expressions in
the long time limit. 
Plots of $\Omega_h$ for $\Delta t_1=\Delta
t_2=\tau$ (\ref{cheq31}), are shown in Fig 3. Very
similar dependence on the model parameters is shown between $\Omega_h$
and $\bar{\epsilon}$, in particular the anomalous $s_1$ behaviour.
\begin{center}
(a)\hspace{6cm}(b)\\
\graphics{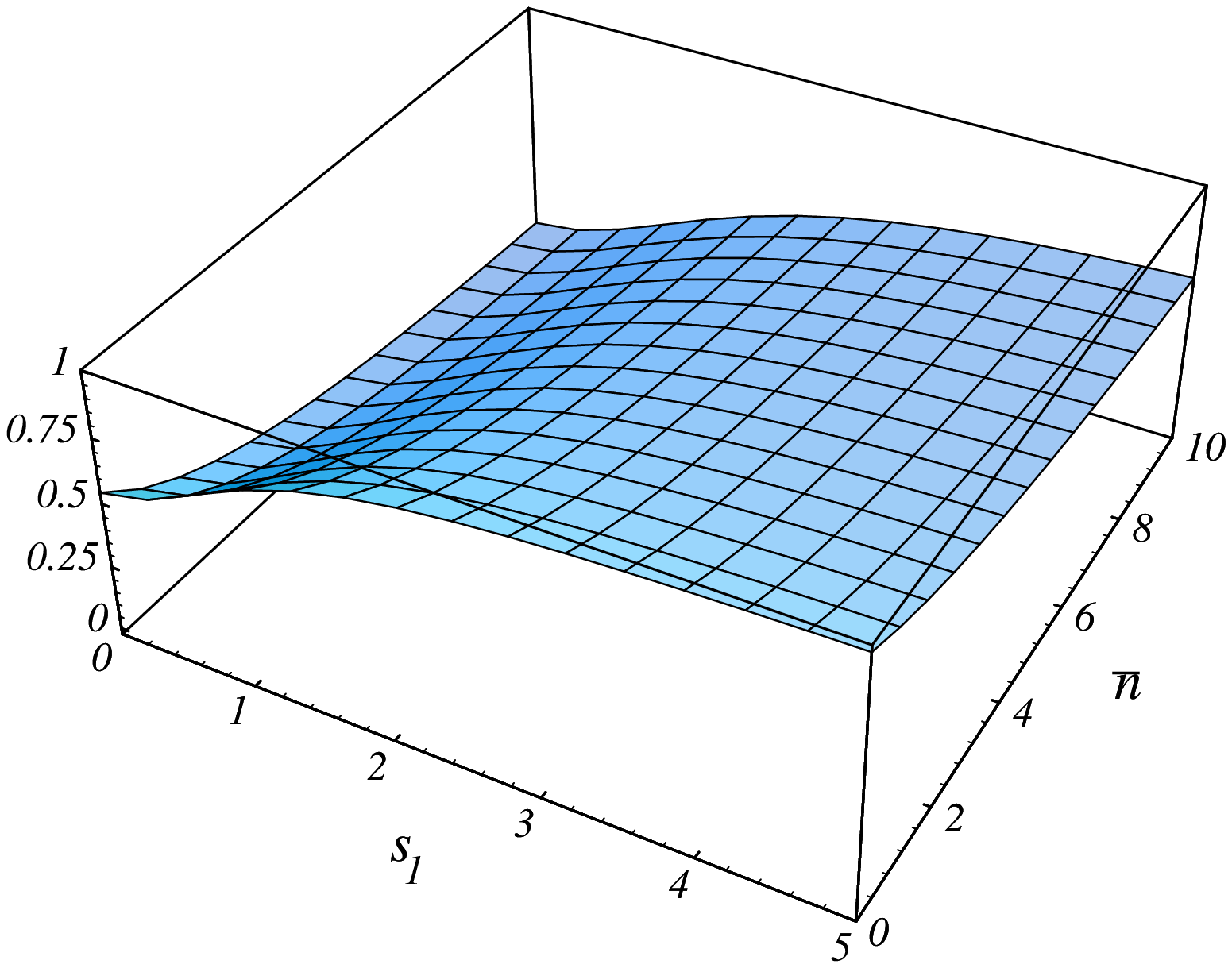}{5}{7}{-2}{-8}{.4}
\graphics{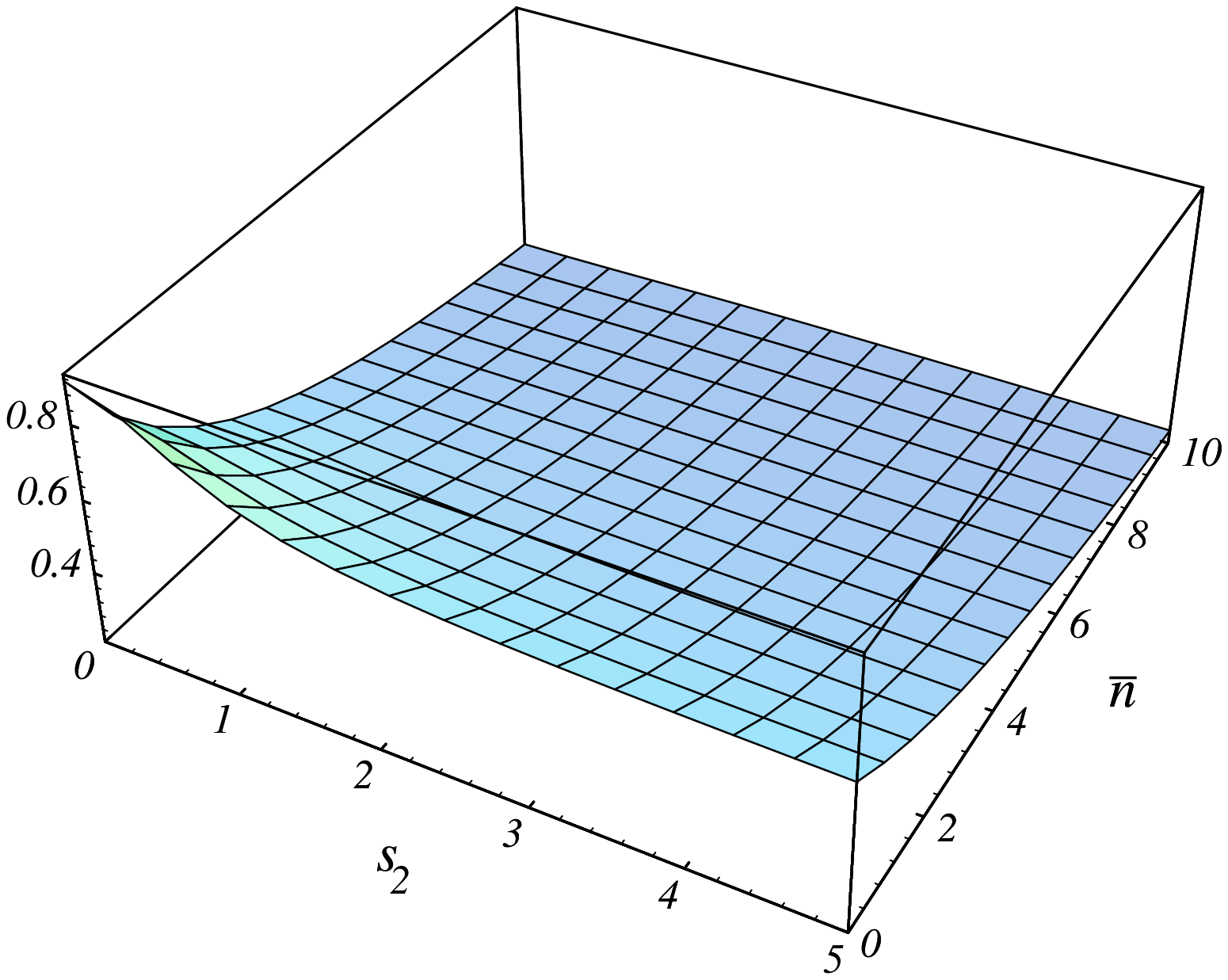}{5}{7}{1}{-8}{.4}
\end{center}
\begin{center}
{\parbox{15cm}{\small  Fig 3.
Plots of $\Omega_h$, (\ref{cheq31}) as a function of $\bar{n}$, $s_1$ and
$s_2$. We have set $\gamma=1$ and $\Delta t_1=\Delta t_2=\tau =0.2\;$. 
Graph (a) 
shows that consistency decreases with increasing initial grain size $s_1$. 
}}
\end{center}

To conclude and summarize: we have examined the Consistent Histories framework
within a quantum optical model using two-time projected phase-space
histories. In the Markovian regime the bath can be traced away and
it's effect completely absorbed into the non-unitary evolution
super-operator $\cal L$. This is used to evolve the reduced density
matrix of the system between projections. Since no true phase-space
projectors exist we use ``quasi-projectors'' which are gaussian
convolutions of coherent states. Although these quasi-projectors are
complete, they are not totally exclusive and therefore do not
exactly meet the specifications of this paradigm. They, however, do
provide a tractable model and have been used successfully in other
calculations. Moreover, such quasi-projectors arise naturally in 
more rigorous theories of quantum measurement. We first calculated the
``action'' of the decoherence functional for these coarse grained
histories on the state $|\alpha_j\rangle\langle\alpha_i|$ at $t_1$.
With this information, the action of the decoherence functional on any
state at $t_1$ can be evaluated through a decomposition of $\rho(t_1)$
into coherent states. This was done for the particular case of
$\rho(t_0)=|\alpha_0\rangle\langle\alpha_0|$ where $t_0 \leq t_1$. The
resulting decoherence functional was exponential in form and peaked on the
classical path of the damped oscillator. Using  the measures of Dowker
and Halliwell the degree of consistency and the degree of peaking about
the classical path were calculated and the short and long time limits
found. The behaviour of these quantities were mostly as expected: the
consistency gets better with time, increasing bath temperature and final
grain size while the peaking is better with increased final and
initial grain size and small bath temperature. The consistency {\em
decreased} with increased initial grain size. We also calculated $\Omega_h$, a
measure similar to the one proposed in the EIS section of this paper,
which quantifies the amount of coherence in the decoherence functional
when the decoherence functional is considered as a density matrix on
the space of coarse grained histories. This measure displayed similar
behaviour to the measure of consistency as advanced by Dowker and
Halliwell.

\section{Conclusion}
Upon comparison we see that both EIS and consistency among phase-space
histories improves as the bath temperature increases. Altering the
coupling strength $\gamma$, changes only the time derivatives of the 
degree of EIS or
consistency achieved and not their actual values. A time for
``decoherence'' may be set equal to the time needed for $\Omega_\eta$ to
reach some predetermined value ie. $\Omega_\eta (t_d)=.5$, say.
The anomalous dependence of $\bar{\epsilon}$ on $s_1$ occurs in all
the two-time projected history calculations to date. For this to occur
in one model for a special type of graining might be understandable
within the arguments given by Dowker and Halliwell concerning the
constructive interference of the constituent histories in a
very special coarse-graining. Since the behaviour is seen in three
different calculations with two different models
\cite{DOWKER:1992,PAZ:1992a} one suspects that there is some feature
common to all that is upsetting the expected behaviour of increased
consistency with increased $s_1$. An approximation common to all of
these calculations is the use of ``quasi-projectors''. It may be that
the non-exclusivity of these projectors is somehow overcounting the
contributions of ``off-diagonal'' histories in the decoherence
functional. It is thus necessary to resort to true projectors. Since
there do not exist true projectors on phase-space one must consider
true projectors on to other spaces. An obvious choice is either the
position, momentum or number basis. Choosing the latter is the most
computationally efficient and was treated in \cite{TWAMLEY:1993b}. The
results were very surprising and highlighted the differences between
the two paradigms, EIS and CH. 

Essentially the goal of EIS is the damping of the coherences present
in the system's quantum state. The goal of CH, however, is the
attainment of histories whose probabilities satisfy the sum rule. For
this to occur the quantum {\em interferences} between the individual
histories must vanish. The damping of quantum coherence and vanishing
of quantum interferences need not be synonymous. The decay of
coherence necessitates the vanishing of interference but not visa
versa. This is clearly shown in the double slit examples of Gell-Mann
and Hartle. When the slit separation is much greater than the
slit/screen separation the resulting coarse-grained histories are very
consistent. However, if the separate beams are then later on
recombined somehow, they will display interference. One then says
that the histories which include this recombination are no longer
consistent. Thus, histories which may begin very consistent, may,
later on, become very inconsistent and visa versa. This is so, even
though the evolution can be strictly unitary! From the viewpoint of
EIS, the coherence of the state is not lost under unitary evolution
and thus EIS does not occur. However, if one now inserts the Young's
slits in a bath of electrons \cite{HARTLE:1989}, the resulting
histories are even more consistent and furthermore they cannot
usually be made less consistent through recombination etc. This is so
because the reduced state of the light beams has lost some of it's
coherence to the bath of electrons. Thus, both CH and EIS has occurred in the
reduced state of the light beams. Since the measure $\Omega_\eta$ was
introduced in section III as a measure of coherence, its value as a measure of
consistency (which involves the vanishing of interferences) 
is slightly suspect. However, we would argue that 
the vanishing of interference between
{\em all} pairs of possible histories in a set of very many alternatives
necessitates the vanishing of interference between neighboring
histories. For this to happen, as in the Young Slit case for small
slit separation, one must have a loss of coherence (usually by
introducing a bath). 

Finally, it would be of greater interest to have a more information
theoretic formulation of the concept of ``diagonalisation'' and
consistency. The introduction of a bath causes the reduced quantum
state to lose it's non-local correlations preferentially in certain
bases. To link this concept with that of a pointer basis - that basis
least effected through the bath interaction - would be of interest.
Also to see how the decohered state (either via EIS or CH) satisfies
the Bell inequalities would give, perhaps, a more physical measure of
what ``diagonality'' really means. This work is in progress 
\cite{TWAMLEY:1993d}. 
Also, the quantification of classical correlations via information
theory in the CH approach is also important \cite{HALLIWELL:1993}.

In this paper we have shown that in this relatively simple model of a
quantum open system EIS and CH seem to behave similarly, becoming
better with increasing bath temperature and time. The dependence of the
consistency on graining also behaves almost as expected. This example
shows how the two paradigms are similar.
Other calculations also demonstrate the differences between the two
\cite{TWAMLEY:1993b,LAFLAMME:1993}. Clearly, if both paradigms are to
describe an ontology it is essential to discover situations where
they differ and to experimentally determine which is correct.

\acknowledgments
The author thanks 
F. Dowker, 
S. Habib,
B. L. Hu 
R. Laflamme, 
A. Matacz, 
J. McCarthy, 
G. Milburn,
J. P. Paz, 
M. Simpson,
W. Unruh and
W. Zurek
for interesting discussions. We also thank the University of
British Colombia, Los Alamos National Laboratories and the University
of Maryland for their kind hospitality during the course of this research.

\appendix
\section{Integral Relations for Coherent States}
\widetext
In this appendix we compile a number of integral results useful for
calculations involving coherent states. We refer the reader to
\cite{ZHANG:1990} for a very complete review of coherent states and their
properties. Here we list a number of results.
\begin{eqnarray}
\int\frac{d^2\alpha}{\pi}\,\exp
&&\left\{
-A|\alpha|^2+B\alpha^2+C\alpha^{*\,2}+D\alpha+E\alpha^*\right\}=\nonumber\\
&&\frac{1}{\sqrt{A^2-4BC}}\exp
\left\{\frac{DEA+E^2B+D^2C}{A^2-4BC}\right\}\qquad\qquad
\left\{\begin{array}{l} A-B-C>0 \\ A^2-4BC>0 \end{array}\right.\;\;,
\label{appendix1}\end{eqnarray}
\begin{eqnarray}
\int\frac{d^2\alpha}{\pi}\,\{\alpha,\alpha^{*}\}\,e^{-A|\alpha|^2
+B\alpha^*+C\alpha}&=&\frac{\{B,C\}}{A^2}\,e^{BC/A}\;\;,\\
\int\frac{d^2\alpha}{\pi}\,\{\alpha^2,\alpha^{*\,2}\}\,e^{-A|\alpha|^2
+B\alpha^*+C\alpha}&=&\frac{\{B^2,C^2\}}{A^3}\,e^{BC/A}\;\;,\\
\int\frac{d^2\alpha}{\pi}\,|\alpha|^2\,e^{-A|\alpha|^2+B\alpha^*+C\alpha}&=&
\frac{1}{A^2}\left[1+\frac{BC}{A}\right]\,e^{BC/A}\;\;,\\
\int\frac{d^2\alpha}{\pi}\,|\alpha|^4\,e^{-A|\alpha|^2+B\alpha^*+C\alpha}&=&
\frac{1}{A^5}\left[2A^2+4ABC+B^2C^2\right]\,e^{BC/A}\;\;.\\
\end{eqnarray}
\begin{eqnarray}
\int\frac{d^2\alpha_1d^2\alpha_2}{\pi}\,
\exp &&\left\{-A_1|\alpha_1|^2+\alpha_1^*[B_1+C_1\alpha_2]+D_1\alpha_1
+E_1\alpha_1^{*\,2}+F_1\alpha_1^2\right.\nonumber\\
&&\left.-A_2|\alpha_2|^2+C_2\alpha_2^*+D_2\alpha_2
+E_2\alpha_2^{*\,2}+F_2\alpha_2^2\right\}=
\frac{1}{\sqrt{\eta}}\,\exp(\Lambda/\eta)\;\;,\end{eqnarray}
where
\begin{equation}
\eta=(A_1^2-4E_1F_1)(A_2^2-4E_2F_2)-4E_2F_1C_1^2\;\;,\end{equation}
\begin{eqnarray}
\Lambda= && (A_1^2-4E_1F_1)(A_2C_2D_2+C_2^2F_2+E_2D_2^2)\\
+&&(A_2^2-4E_2F_2)(A_1B_1D_1+B_1^2F_1+E_1D_1^2)\\
+&&(D_1A_1+2B_1F_1)(A_2C_2+2E_2D_2)C_1\\
+&&C_1^2D_1^2E_2+C_1^2C_2^2F_1\;\;.\end{eqnarray}

\section{Position Resolution of $\rho$}
In this appendix we calculate the quantity $\langle x|e^{-\xi
a^\dagger} e^{\xi^{*}a}|y\rangle$. This is used in (\ref{xy1}) to
obtain the resolution of the density matrix in the position basis.
 From the group properties of coherent states and the
Baker-Campbell-Hausdorff expansion we can get
\begin{eqnarray}
\langle x|e^{-\xi
a^\dagger} e^{\xi^{*}a}|y\rangle &=&
\langle x|e^{-\xi
a^\dagger+\xi^{*}a+|\xi|^2/2}|y\rangle\\
&=&e^{|\xi|^2/2}\langle x|\exp\left[
\hat{x}\sqrt{\frac{\omega}{2\hbar}}(\xi^*-\xi)
+i\frac{\hat{p}}{\hbar}\sqrt{\frac{\hbar}{2\omega}}(\xi^*+\xi)\right]|y\rangle
\label{in1}\\
&=&e^{|\xi|^2/2+(\xi^{*\,2}-\xi^2)/4}
\exp\left[x\sqrt{\frac{\omega}{2\hbar}}(\xi^*-\xi)\right]
\langle x|\exp
\left[i\frac{\hat{p}}{\hbar}\sqrt{\frac{\hbar}{2\omega}}(\xi^*+\xi)\right]
|y\rangle\;\;,\label{in2}\end{eqnarray}
where we have re-written the $a$, $a^\dagger$ in terms of $\hat{x}$ and
$\hat{p}$ and again used  BCH identities to dis-entangle
$e^{A\hat{x}+B\hat{p}}= e^{A\hat{x}}e^{B\hat{p}}e^{-AB[x,p]/2}$.
Now since $e^{-i\hat{p}x/\hbar}|y\rangle=|y+x\rangle$ the inner
product on the right hand side of (\ref{in2}) becomes the delta
function
\begin{equation}
\langle x|\exp
\left[i\frac{\hat{p}}{\hbar}\sqrt{\frac{\hbar}{2\omega}}(\xi^*+\xi)\right]
|y\rangle=
\delta(\xi_x+\sqrt{\frac{\omega}{2\hbar}}(x-y))
\sqrt{\frac{\omega}{2\hbar}}\;\;,\end{equation}
where $\xi_x={\rm Re}\,\xi$.
Collecting the terms we finally get,
\begin{equation}
\langle x|e^{-\xi a^\dagger}e^{\xi^* a}|y\rangle=
\exp\left[
\frac{|\xi|^2}{2}+\frac{\xi^{*\,2}-\xi^2}{4}+x\sqrt{\frac{\omega}{2\hbar}}
(\xi^*-\xi)\right]\delta(\xi_x+\sqrt{\frac{\omega}{2\hbar}}(x-y))
\sqrt{\frac{\omega}{2\hbar}}\;\;.\end{equation}
\narrowtext

\end{document}